\documentclass[a4,amsmath,amssymb,superscriptaddress, showpacs, twoside, epsfig]{revtex4}

\usepackage{color}

\usepackage{bbold}

\usepackage{graphicx}% Include figure file
\usepackage{epstopdf} 
\usepackage{caption}
\usepackage{subfig}
\usepackage{pdfsync}
  
\newcommand{\be}[1]{ \begin{eqnarray} \mbox{$\label{#1}$} }

\newcommand{\ee}{\end{eqnarray}}

\newcounter{mycount}

\newcommand{\p}{\partial}

\newcommand\noi{\noindent}
\newcommand{\bi}{\begin{itemize}}
\newcommand{\ei}{\end{itemize}}
\newcommand{\ba}{\begin{eqnarray}}
\newcommand{\ea}{\end{eqnarray}}

\begin{document}

\title{Rotational properties of two-component Bose gases in the lowest Landau level}
\author{M.L. Meyer} 
\affiliation{ Department of Physics, University of Oslo, P.O. Box 1048 Blindern, 0316 Oslo, Norway}

\author{G.J. Sreejith}
\affiliation{Nordita, KTH Royal Institute of Technology and Stockholm University, Roslagstullsbacken 23, SE-106 91 Stockholm, Sweden}
\author{ S. Viefers}
\affiliation{ Department of Physics, University of Oslo, P.O. Box 1048 Blindern, 0316 Oslo, Norway}

\date{\today}

\begin{abstract} 
We study the rotational (yrast) spectra of dilute two-component atomic Bose gases in the low angular momentum regime, assuming equal interspecies and intraspecies interaction. Our analysis employs the composite fermion (CF) approach including a pseudospin degree of freedom. While the CF approach is not {\it a priori} expected to work well in this angular momentum regime, we show that composite fermion diagonalization gives remarkably accurate approximations to low energy states in the spectra. For angular momenta $0 < L < M$ (where $N$ and $M$ denote the numbers of particles of the two species, and $M \geq N$), we find that the CF states span the full Hilbert space and provide a convenient set of basis states which, by construction, are eigenstates of the symmetries of the Hamiltonian. Within this CF basis, we identify a subset of the basis states with the lowest $\Lambda$-level kinetic energy. Diagonalization within this significally smaller subspace constitutes a major computational simplification and provides very close approximations to ground states and a number of low-lying states within each pseudospin and angular momentum channel.
\end{abstract}
\pacs{ }

\maketitle

\section{Introduction}
\label{sec:intro}

In recent years there has been extensive interest in the study of strongly correlated states of cold atoms motivated by analogies with exotic states known from low-dimensional electronic systems. Substantial theoretical and experimental effort is being devoted to the possible realisation of quantum Hall-type states in atomic Bose condensates \cite{reviews}. Conceptually the simplest way of simulating the  magnetic field is by rotation of the atomic cloud  \cite{roncaglia11}, although alternative proposals involving synthetic gauge fields \cite{lin09, dalibardreview, julia-diaz} are more likely to provide stronger magnetic fields. Consequently, a large body of work has focused on the rotational properties of atomic Bose gases. In this context, several groups have studied the rotational properties of bosons in the lowest Landau level also at the lowest angular momenta \cite{mottelson99, bertsch99, smith00, kavoulak01, korslund, viefers10} ($L \leq N$ where $N$ is the number of particles). Notably, analytically exact ground state wave functions were found in this regime\cite{bertsch99, smith00}. Much of these studies have considered small systems. Interestingly, a recent experimental paper \cite{gemelke10}, claiming the first ever realisation of rotating bosons in the quantum Hall regime, precisely involves such small systems (up to ten particles), and includes the lowest angular momenta.

Even richer physics can be expected in the case of {\it two-species} bosons (or fermions, for that sake). Two-species Bose gases can be realised as mixtures of two different atoms\cite{mondugno02}, two isotopes of the same atom\cite{bloch01}, or two hyperfine states of the same atom\cite{hall98}. As long as all interactions, between atoms of the same species, and between atoms of different species, are the same (we will refer to this as 	`homogeneous interaction'), the system possesses a pseudospin symmetry, with the two species corresponding to pseudospin ``up" and ``down", respectively. Tuning the interaction away from homogeneous can lead to interesting physics, such as a transition from a miscible to an immiscible regime where the interspecies interaction dominates\cite{papp08}. It is then obviously of interest to study the rotational properties of such systems. Several recent papers have addressed the very interesting topic of possible quantum Hall phases of two-species Bose gases \cite{grass12, ueda12, jain13, furukawa13, senthil13, grass13}. 

In this paper we study the yrast spectra of dilute, rotating two-species Bose gases with homogeneous interaction, at low angular momenta, $L\leq N+M$, where $N+M$ is the total number of particles. Physically, this is the regime where the first vortices (in the two components) enter the system\cite{bargi07}. A recent study by Papenbrock et al\cite{papenbrock} identified a class of analytically exact states in the yrast spectrum of this regime; these include the ground state and some (but not all) excited states. Our approach is to study the yrast spectrum in terms of trial wave functions given by the composite fermion (CF) approach\cite{jainbook}, with a pseudospin degree of freedom accounting for the two species. As has been discussed for the one-species case \cite{viefers00, korslund, viefers10}, this is a regime where CF trial wave functions cannot, {\it a priori}, be expected to work well, since the physical reasoning behind the CF approach \cite{jainbook} implies that it should apply primarily in the quantum Hall regime where angular momenta are much higher, $L \sim N^2$. Nevertheless, this approach turned out to work very well for single species gases, and as we will see, the same happens for two-species systems.

We present results for up to twelve particles, in the disk geometry \cite{grasscomment}. We find that the CF formalism gives very good approximations to the exact low energy states of the system, with overlaps very close to unity. In particular, all states of the yrast spectrum, including those of Ref. \onlinecite{papenbrock}, at $0 < L < M$ are reproduced {\it exactly}.
 
For given $N,M$ and $L$, the number of CF candidate states of the right quantum numbers is usually larger than one, and one performs a diagonalization within the space spanned by these states. This is what we will refer to as CF-diagonalization. However, the dimension of this CF subspace is often not significantly smaller than the dimension of the full Hilbert space, and therefore the computational gain over a full numerical diagonalization may not be significant. We have identified, within the basis sets of CF candidates for given $(N,M,L)$, a class of states of particularly simple structure, which in themselves have almost complete overlap with the lowest lying states. These states (which we will refer to as `simple states'), are characterised by having at most one composite fermion in each $\Lambda$-level, as will be explained below. Because of the pseudospin symmetry provided by the homogeneous interaction, from some state at given $L$ and $S_z = (M-N)/2$, one can obtain, by pseudospin lowering, a state which has the same energy (and overlap) at different $M$ and $N$. Typically, this is an excited state at the new $M, N$. In this way, the simple states will produce excellent trial wave functions for the ground states, as well as a number of low-lying states for the whole pseudospin multiplet at given total number of particles $N+M$. Restriction to the simple CF states constitutes a major computational simplification due to reduction in the dimension of the space. Thus, one gets, with relative ease, access to explicit polynomial wave functions that are very close to exact and can be further used to study, e.g., the vortices in the system.

The paper is organised as follows. In section \ref{sec:theory} we introduce the theoretical background and formalism used. Results for full CF diagonalization as well as simple state calculations are presented in section \ref{sec:results}. Finally, in section \ref{sec:concl}, we summarise and discuss future perspectives.

\section{Theory and methods}
\label{sec:theory}

\noindent
{\it System and observables:}
The system we study consists of two species of bosons in a two dimensional rotating harmonic trap, interacting with each other through a contact potential. We will use the pseudospin terminology and refer to one species as "up" ($\uparrow$) and the other as "down" ($\downarrow$). The system of $N$ bosons of type $\downarrow$ and $M \geq N$ bosons of type $\uparrow$ is described by the Hamiltonian
\begin{equation}
H = \sum_{i=1}^{N+M} \left( \frac{\mathbf{p}_i^2}{2m}+\frac{1}{2}m\omega^2\mathbf{r}_i^2 - \Omega l_i \right)
   + \sum_{i,j=1}^{N+M}2\pi g \delta(\mathbf{r}_i - \mathbf{r}_j)
\end{equation}
where $m$ is the mass of the bosons, $\omega$ is the harmonic trap frequency, $\Omega$ is the rotational frequency of the trap around the $z$-axis, $l_i$ are the one-body angular momenta in the $z$-direction, and $g$ is a parameter specifying the two-body interaction strength.
As usual \cite{reviews}, in the dilute (weakly interacting) limit, this model may be recast as a two-dimensional lowest Landau level problem in the effective magnetic field $B_{eff}=2m\omega$,
\begin{align}
H &= \sum_i^{N+M}(\omega-\Omega)l_i + 2\pi g \left[ \sum_{i<j=1}^{N} \delta(z_i - z_j) +\dots\right.\nonumber\\
&\qquad\qquad \left. \dots+ \sum_{k<l=1}^{M} \delta(w_k - w_l) + \sum_{i<k=1} \delta(z_i - w_k) \right]
\end{align}
with 'flat' Landau levels corresponding to the ideal limit $(\omega - \Omega) \rightarrow 0$. Here $z = x+iy$ now refers to the positions of the $\downarrow$-particles, and $w$ refers to the $\uparrow$-particles. Notice that we have assumed equal interspecies and intraspecies interaction strength, as well as equal masses for the two species.

The single-particle states spanning the lowest Landau level (in the symmetric gauge) are
\begin{equation}
\psi_{0,l}(z) = N_{l} z^l \exp{(-z\bar{z}/4)} \qquad l\geq 0 \label{spwf_lll}
\end{equation}
where $l$ is the angular momentum of the state and $N_l$ is a normalisation factor. The unit of length is $\sqrt{\hbar/(2m\omega)}$. Hereafter we will omit the ubiquitous Gaussians and assume that derivatives do not act on the Gaussian part of the wavefunctions. The many-body wavefunctions with total angular momentum $L$ are then homogeneous symmetric (separately in $z_i$ and $w_k$) polynomials of total degree $L$ in the coordinates $z_i, w_k$.

The Hamiltonian is symmetric under rotation and under change of species of bosons (pseudospin). The primary observables of interest (that commute with the Hamiltonian) in the following discussion are the total angular momentum 
\begin{equation}
L_z = \sum_{i=1}^N z_i\partial_{z_i} + \sum_{k=1}^M w_k\partial_{w_k},
\end{equation}
the center of mass angular momentum
\begin{equation}
L_c = R\left( \sum_{i=1}^N \partial_{z_i} + \sum_{k=1}^M \partial_{w_k} \right), \label{LC}
\end{equation}
where $R=\frac{1}{N+M}\left( \sum_i^N z_i + \sum_k^M w_k \right)$ is the center of mass coordinate, and finally
the pseudospin-1/2 operators $\mathcal{S}^2$ and $\mathcal{S}_z$ which, in second quantization language, are defined as 
\begin{equation}
\mathcal{S}_z = \frac{1}{2}\sum_{l=0}^\infty b_{\uparrow,l}^\dagger b_{\uparrow,l} - b_{\downarrow,l}^\dagger b_{\downarrow,l}
\end{equation}
\begin{equation}
\mathcal{S}^2 = \mathcal{S}_{-}\mathcal{S}_{-}^\dagger + \mathcal{S}_z(\mathcal{S}_z+1).
\end{equation}
Here
\begin{equation}
\mathcal{S}_{-} = \sum_{l=0}^\infty b_{\downarrow,l}^\dagger b_{\uparrow,l}
\end{equation}
is the pseudospin lowering operator, taking a state at $(N,M)$ to a state at $(N+1,M-1)$. The $b_{\downarrow,l}^\dagger$ and $b_{\uparrow,l}$ are creation and annihilation operators for $\downarrow$-type and $\uparrow$-type bosons at angular momentum $l$ respectively.

As stated before, the eigenfunctions of $L_z$ are homogeneous, symmetric polynomials of total degree $L$. The eigenvalues of $L_c$ are $l_c = 0,1,\ldots,L$ and the corresponding eigenfunctions are given by $R^{l_c}\Psi(z,w)$. The function $\Psi(z,w)$ satisfies
\begin{equation}
 L_c\Psi(z,w) = 0
\end{equation}
and is therefore a translationally invariant (TI) state (see Appendix \ref{appA})
\begin{equation}
 \Psi(z-c,w-c) = \Psi(z,w).
\end{equation}
We will focus on such states in this paper, realizing that we can create all the states of energy $E$, angular momentum $L+k$ and center-of-mass (COM) angular momentum $k$ (called COM excitations) by multiplication of $R^k$ with wave functions of energy $E$, angular momentum $L$ and zero COM angular momentum.

For a given $N$ and $M$, the eigenvalue of $\mathcal{S}_z$ is $(M-N)/2$, while the eigenvalues of $\mathcal{S}^2$ are  $S(S+1)$, with $S=S_z,S_z+1,\ldots,(N+M)/2$. Since $\left[ \mathcal{S}^2,\mathcal{S}_z \right] =0$, we can
find the eigenstates in the following way: for a given $A=N+M$, we find energy eigenstates at $N=0,M=A$. These states all have $S=S_z=(M-N)/2=A/2$, the only possible value for $S$.
Applying the pseudospin lowering operator to these states, we get all possible states at $S=A/2$ for the case $N=1,M=A-1$.
The remaining states at $N=1,M=A-1$ must then be the ones with $S=S_z=(A-2)/2$. 

Since the pseudospin operators commute with the interaction, both energy expectation values and overlaps are unchanged under pseudospin raising and lowering. Therefore, for the purpose of comparing the exact energy eigenstates with the CF states of the same quantum number, we consider only the translationally invariant ($L_c=0$) highest weight states ($S=S_z$), abbreviated as TI-HW states in the following.

\bigskip
\noindent
{\it Composite fermion trial wave functions:}
The composite fermion (CF) approach \cite{jainbook} has been extremely successful in describing the strongly interacting electrons in quantum Hall systems in terms of composite objects comprised of an electron and an even number of vortices (or, loosely speaking, magnetic flux quanta), moving in a reduced effective magnetic field. The single particle states available to these emergent particles, form Landau-like levels (often called $\Lambda$ levels) in this effective magnetic field. The single particle wavefunction with an angular momentum $m$ in the $n^{\rm th}$ $\Lambda$-level is 
\begin{equation}
\psi_{n,m}(z) = N_{n,m} z^m L_n^m\left( \frac{z\bar{z}}{2} \right),
\quad m \geq -n \label{spwf},
\end{equation}
where $L_n^m$ is the associated Laguerre polynomial, and $N_{n,m}$ is a normalization factor.
As it turns out, composite fermions can be considered as weakly interacting, and in fact, a very accurate description can be obtained by simply approximating them as non-interacting. In this approximation, the generic form of a CF trial wave function is
\begin{equation}
 \Psi_{CF}=\mathcal{P}_{LLL} \left(\Phi \, J^p \right)
\end{equation}
where $\Phi$ is a Slater determinant of CFs,  $J$ is a Jastrow factor, $J = \prod_{i<j}(z_i-z_j)$, so $J^p$ represents the "attachment" of $p$ vortices to each electron ($p$ is an even integer). $\mathcal{P}_{LLL}$ denotes lowest Landau level projection. This approach can be straightforwardly modified \cite{reviews} to describe quantum Hall-type states of {\it bosons}, by letting $p$ be an odd number. Throughout this paper we will work with the simplest case, $p=1$.

The above CF formalism can be generalized \cite{jainbook} to take into account a spin or pseudospin degree of freedom (for example, in a bilayer quantum Hall system). In this work we will pursue the analogy to spinful composite fermions, to study the case of two-species rotating bosons in the lowest Landau level, with a pseudospin-$1/2$ degree of freedom accounting for the two species of particles.
In this case, a CF trial wave function has the general form
\begin{equation}
 \Psi_{CF}=\mathcal{P}_{LLL} \left(\Phi_\downarrow \Phi_\uparrow J(z,w) \right)
 \label{2swf}
\end{equation}
where $\Phi_\downarrow$, $\Phi_\uparrow$ are Slater determinants for each species of the non-interacting CFs (consisting of the single-particle states  (\ref{spwf})), and the Jastrow factor involves both species,
\begin{equation}
J(z,w)=\prod_{i<j=1}^N(z_i-z_j)\prod_{k<l=1}^M(w_k-w_l)\prod_{i,k=1}^{N,M}(z_i-w_k).
\end{equation}
Projection to the lowest Landau level is achieved by replacing the conjugate variables $\overline{z}_i$, $\overline{w}_k$ by $\partial_{z_i}$, $\partial_{w_k}$ after moving them all the way to the left in the final polynomial \cite{jainbook}. 

A technical comment that is in order here concerns the fact that we are studying {\it low} angular momentum states. In the quantum Hall regime ($L\sim \mathcal{O}(N^2)$), where the CF approach is usually applied, many-body ground states involve a small number of filled $\Lambda$ levels. For example, the $\nu=1/3$ Laughlin state is represented as an integer quantum Hall state of composite fermions with one filled $\Lambda$ level, the $\nu=2/5$ quantum Hall state corresponds to two filled $\Lambda$ levels etc. The situation in a low-$L$ state is somewhat different - since the Jastrow factor in (\ref{2swf}) itself contributes a large angular momentum $L_J=A(A+1)/2$ where $A=N+M$, the angular momenta of the Slater determinants have to be {\it negative}. The Slater determinants attain angular momentum by including many derivatives through occupying many $\Lambda$ levels.  For this reason, the CF states in the context of lower angular momenta involve {\it many} $\Lambda$ levels, with only a few CFs in each (see e.g. the $L=N$ CF candidate discussed in Ref. \onlinecite{korslund}). 

\begin{figure}
\includegraphics[width=\columnwidth]{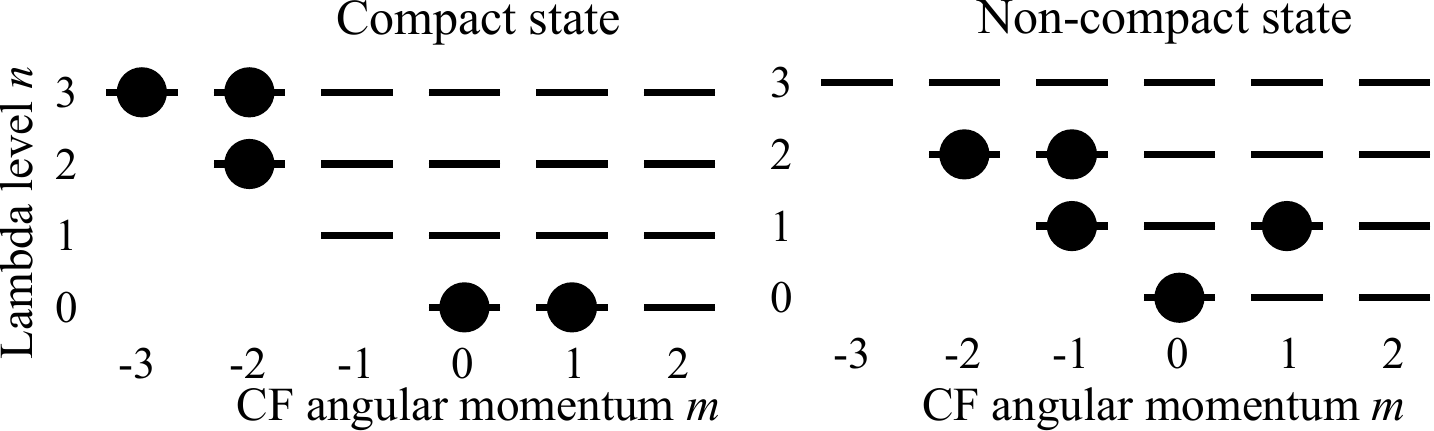}
\caption{Example of compact and non-compact states of a single species shown schematically. The lines indicate the possible single states that can be occupied by the CFs and the dots indicate occupied states\label{fig:compactnoncompact}}
\end{figure}

%
%\bigskip
%\noindent
%{\it Compact CF states:}
The fact that the interaction is translationally invariant and commutes with the center-of-mass angular momentum allows us to focus our attention on ``compact CF states''. A CF state is ``compact" if, for every single particle state $\psi_{n,m}$ in its Slater determinant, the states $\psi_{n-1,m}$ and $\psi_{n,m-1}$ are also present. That is, every occupied $\Lambda$ level in the determinant is occupied ``compactly'', with all states $m=-n,\ldots$ up to the largest value of $m$ in that level occupied. Fig:\ref{fig:compactnoncompact} shows examples of such compact and non-compact states.

All compact states are translationally invariant, and it can be shown that the projected Slater determinant is equivalent to a determinant in which the single particle states have the simple form \cite{jainbook}
\begin{equation}
\psi_{n,m}(x_i) \propto x_i^{n+m}\partial_i^{n}.
\end{equation}
Because we are interested in the highest weight pseudospin states, the CF states that we consider are
required to satisfy Fock's cyclic condition \cite{jainbook}. For the CF states which are simple products of two Slater determinants times a Jastrow factor, Fock's cyclic condition is satisfied iff for every single CF state that is occupied in 
the $\downarrow$ species (the species with fewer particles) the corresponding CF state is occupied in the $\uparrow$ species (majority species) also. When $N=M$, the two Slater determinants contain the same sets of occupied states $(n,m)$. Note that a CF state which is not a simple product of two determinants and a Jastrow factor may also be a highest weight-state; we do not include such states in our analysis. 

By 'full CF-diagonalization' we thus mean a diagonalization of the interaction in the space spanned by compact CF states with correct angular momentum that satisfy Fock's cyclic condition (in the sense described above). Note that the number of linearly independent CF states will generally be less than the number of determinant-pairs that satisfy the relevant restrictions. Therefore we reduce the set of projected wave functions to a linearly independent set before diagonalization.
%
%\noindent
%{\it Simple CF states}

\begin{figure}
\includegraphics[width=\columnwidth]{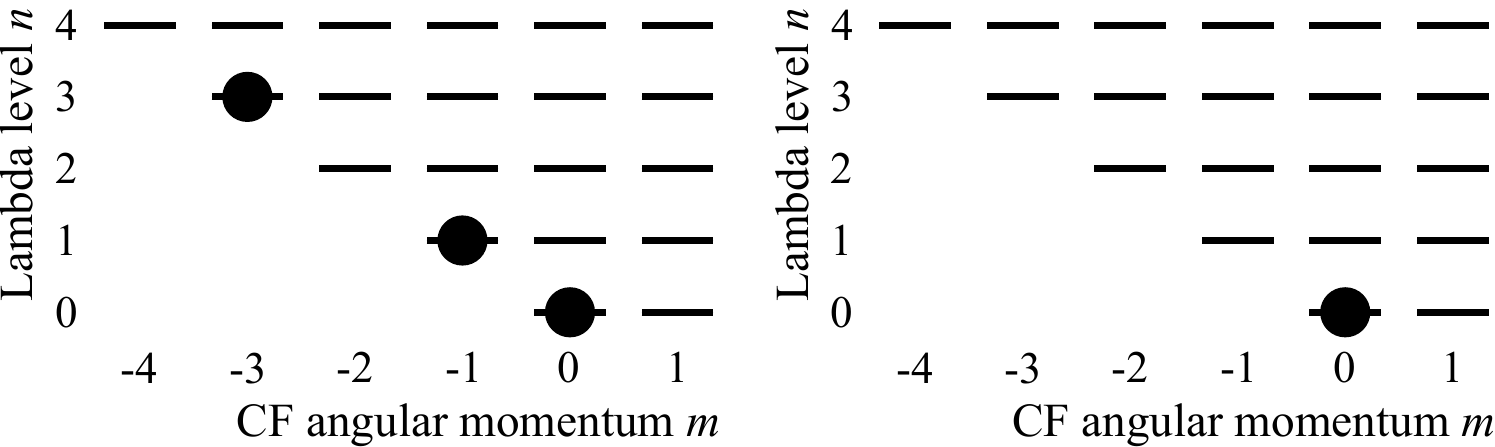}
\caption{Two slater determinants of a simple state shown schematically. Left and right sections of the figure show the distribution of the CFs in the two slater determinants involved in the two boson CF state in Eq \ref{simple_example}. Each $\Lambda$ level contains either one or no CFs.\label{fig:simplestate}}
\end{figure}

A class of CF states to which we will devote special attention is the subset of compact CF states which minimize the total $\Lambda$-level kinetic energy, for a given $N$, $M$ and $L$. As mentioned, many $\Lambda$-levels need to be involved in order to produce the low angular momenta of the projected states we consider. On the other hand, the state vanishes if the $n=A$ level or higher is occupied (due to the high derivatives this implies). For $N>0$ and a given angular momentum $0 < L < N+M$, one observes that the kinetic energy is minimized when the available $\Lambda$-levels are either empty or singly occupied. This means that all the non-zero single particle wave functions in the Slater determinants are
of the form $\psi_{n,-n}$, which in the projected Slater determinants simply become
\begin{equation}
\psi_{n,-n} \propto \partial_i^{n} \label{simple-spwf}
\end{equation}
Such states will be called 'simple states' in this paper. An example of such a state is the CF candidate for $N = 1, M=3, L=2$ in which the single-particle states $(n,m) = (0,0), (1, -1)$ and $(3, -3)$ are occupied by one species; and $(n,m) = (0,0)$ is occupied by the other (see Fig \ref{fig:simplestate}).

%The diagram
%for one of the determinants would be
%\begin{equation}
%\begin{array}{ccccccccc}
% & 4 & \blacklozenge & - & - & - & - & - & -\\
% & 3 &  & - & - & - & - & - & -\\
% & 2 &  &  & \blacklozenge & - & - & - & -\\
%n & 1 &  &  &  & - & - & - & -\\
% & 0 &  &  &  &  & \blacklozenge & - & -\\
% &  & -4 & -3 & -2 & -1 & 0 & 1 & 2\\
% &  &  &  & m\end{array}\label{eq:compact-diagram}
%\end{equation}
The corresponding projected wave function, is 
\be{}
\psi(\{ z_i\}, \{ w_i\}) = 
\begin{vmatrix}
\p^0_{z_1}
\end{vmatrix}
\cdot
\begin{vmatrix}
 \p^0_{w_1} \, \p^0_{w_2} \,\p^0_{w_3}\\
\p^1_{w_1} \, \p^1_{w_2} \,\p^1_{w_3} \\
\p^3_{w_1} \, \p^3_{w_2} \, \p^3_{w_3}  \\
\end{vmatrix}
\cdot
\, J(z,w)
\label{simple_example}
\ee
Another example are the cases $N=M=L$, where the CFs occupy every other $\Lambda$-level, i.e. $(0,0)$, $(2,-2)$ etc. up to $(2L-2,2-2L)$. Incidentally these are exact ground states for the interaction we are considering.

We find that, in order to capture the lowest energy states, it is sufficient to diagonalize within the restricted subspace of simple CF states instead of diagonalizing within all the compact CF candidates.

\section{results}
\label{sec:results}
In this section, we present a comparison of spectra from exact diagonalization of the Hamiltonian in the translationally invariant, highest weight sector to the results from the CF diagonalization discussed in Section II. We will present results both from full CF diagonalization and specifically from the use of only the 'simple states'. Systems of a total of up to 12 particles have been studied. For up to 8 particles, the numerics were done in Mathematica, preserving symbolic (i.e. infinite) precision up to the point where overlaps are given to machine precision. For larger systems a projection algorithm implemented in C was used (Appendix \ref{appB}).

\subsection{Full CF diagonalization}
\begin{figure}[!htb]
  \centering
  \subfloat{\includegraphics[width=0.4\textwidth]{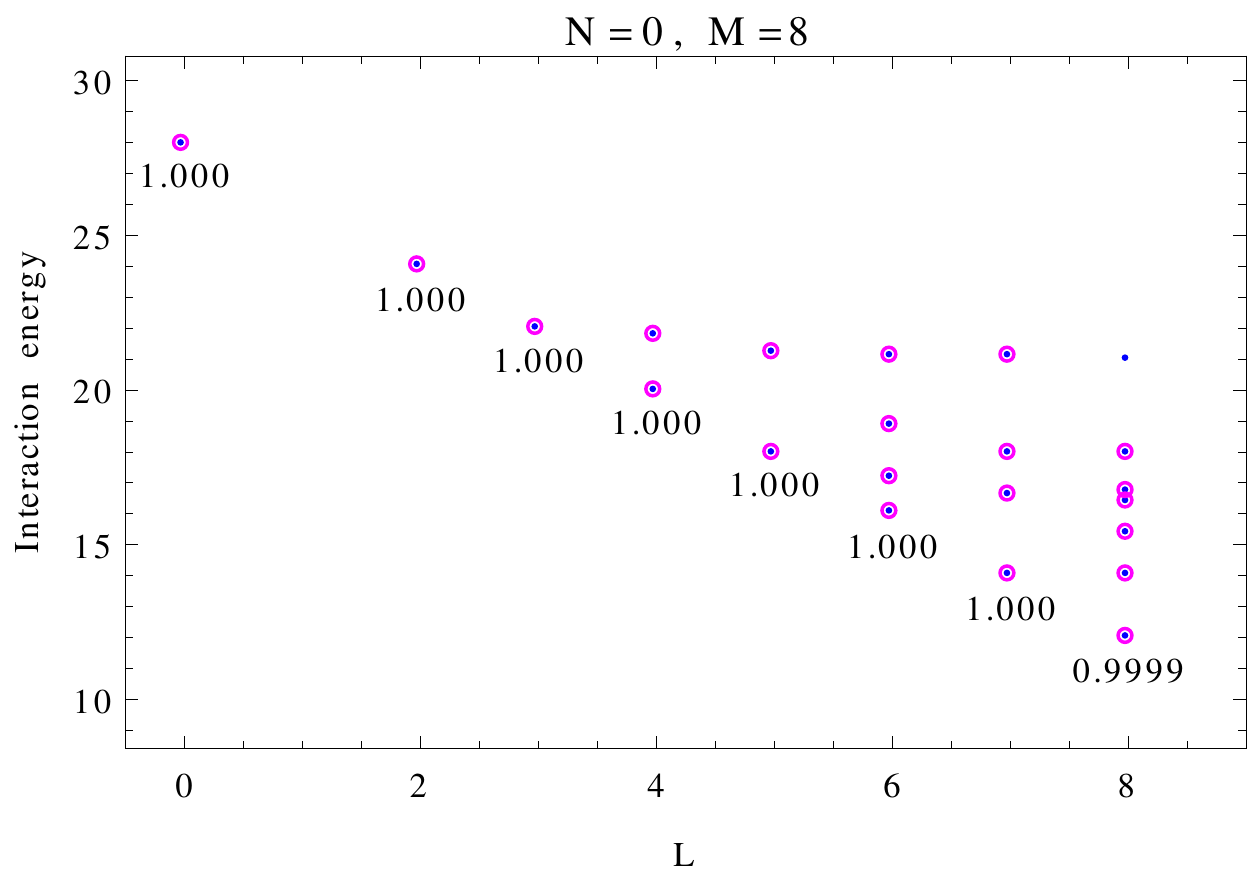}}
  \quad
  \subfloat{\includegraphics[width=0.4\textwidth]{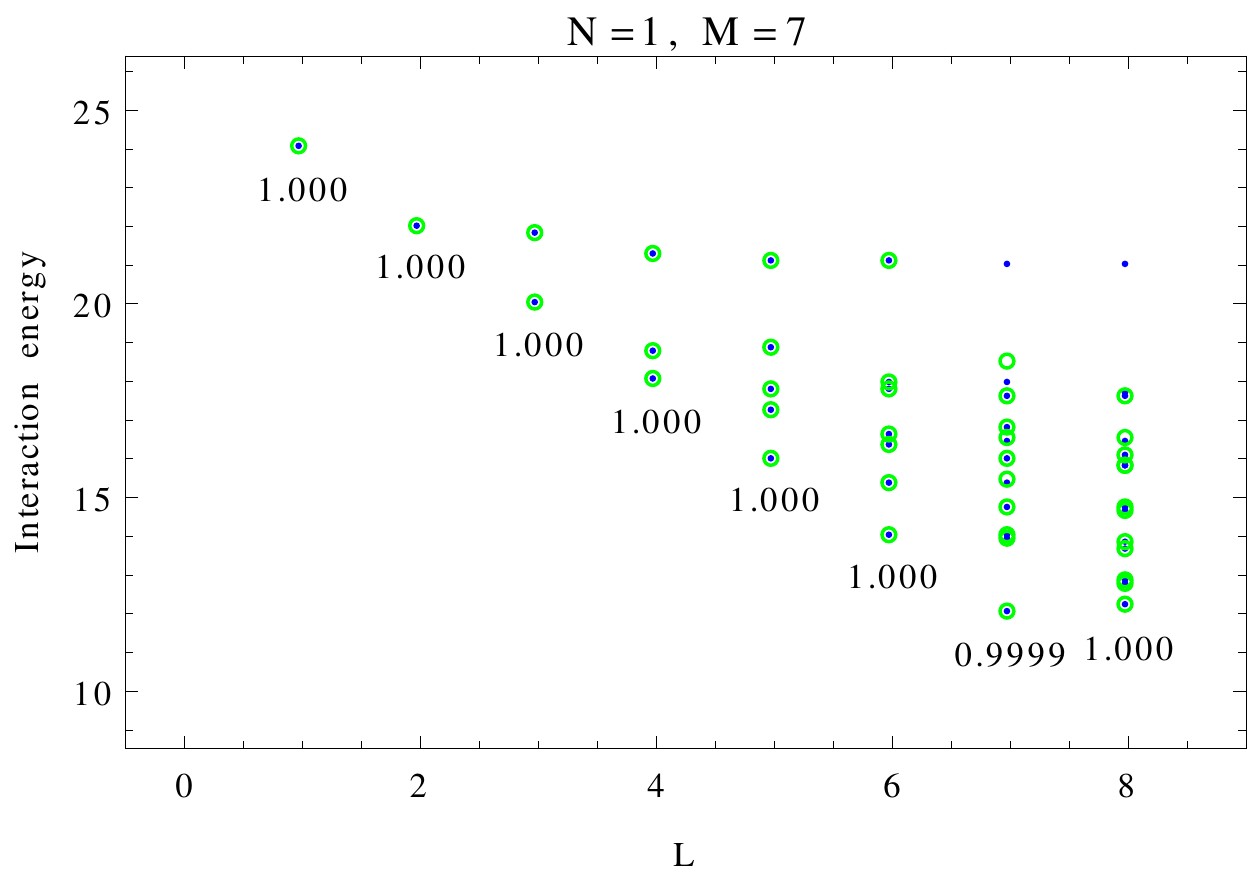}}
  \newline
  \subfloat{\includegraphics[width=0.4\textwidth]{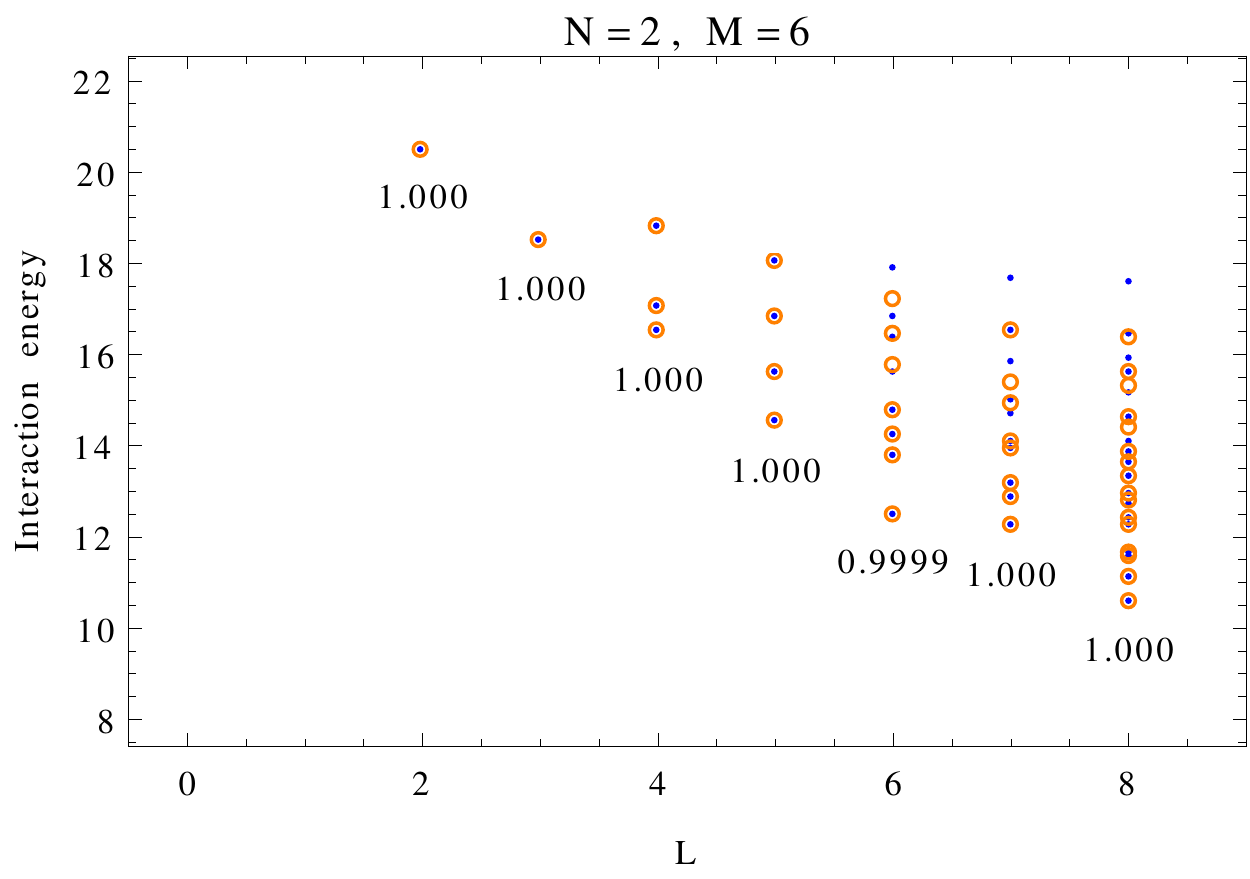}}
  \quad
  \subfloat{\includegraphics[width=0.4\textwidth]{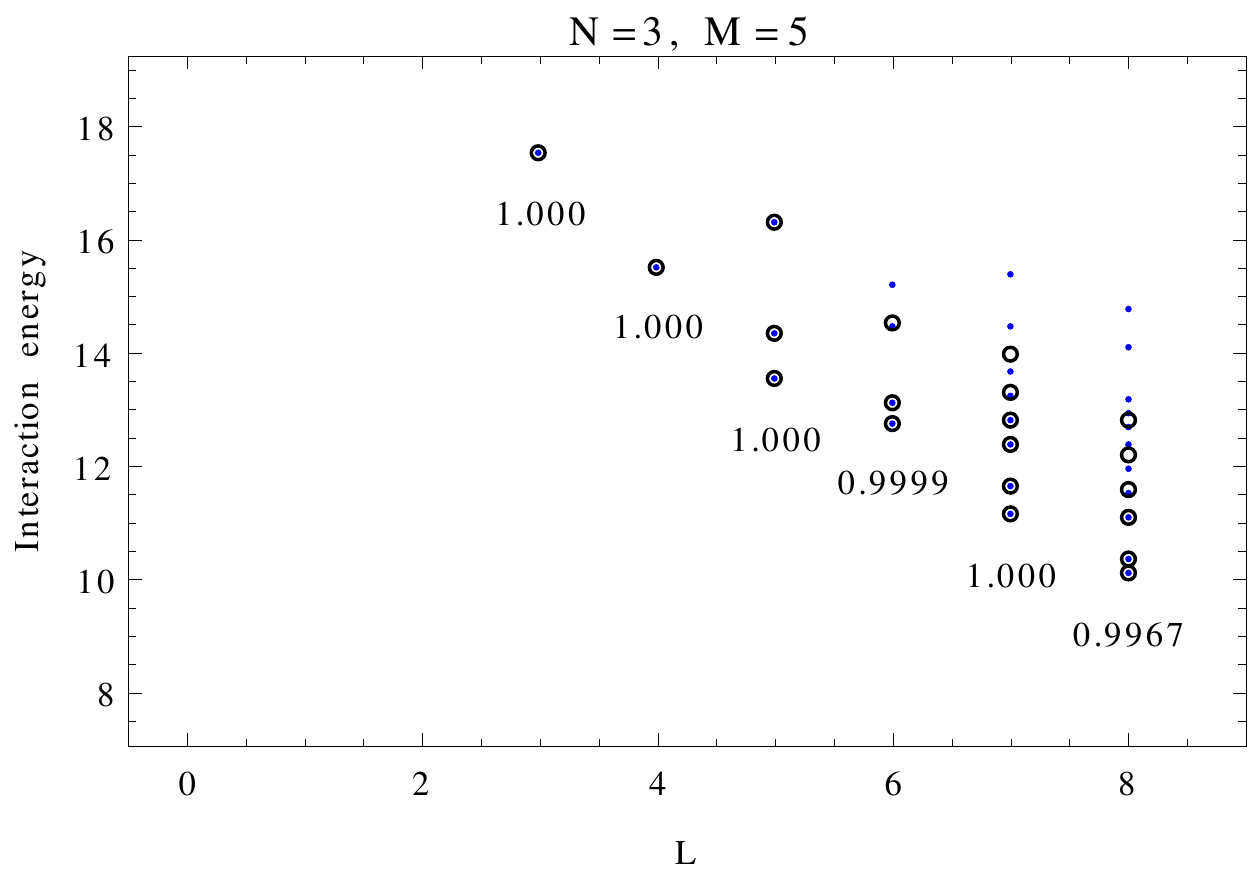}}
  \newline
  \subfloat{\includegraphics[width=0.4\textwidth]{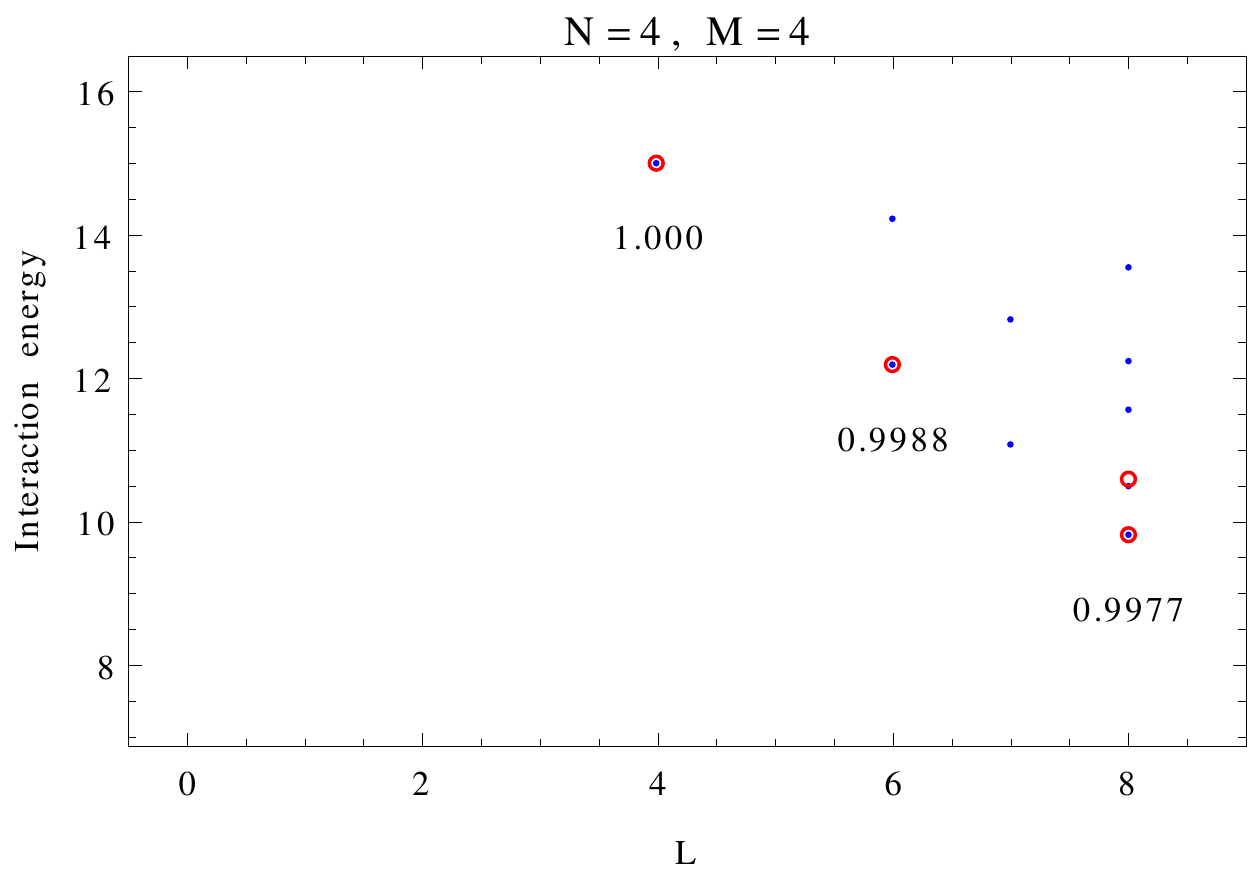}}
  \quad
  \subfloat{\includegraphics[width=0.4\textwidth]{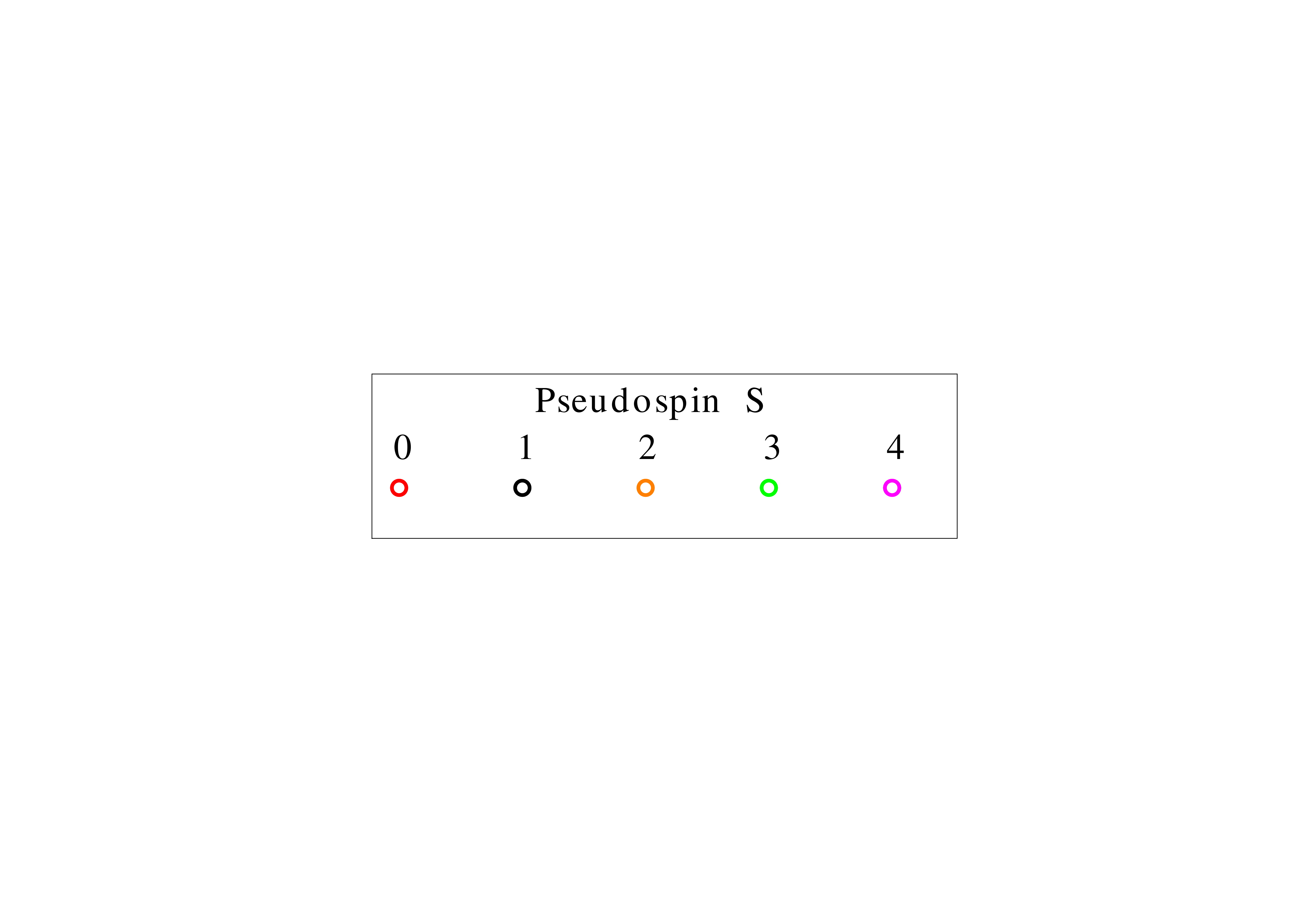}}
  \newline
  \caption{(Color online) Exact energy eigenstates (blue dots) and full CF diagonalization
  results (colored rings) in the TI-HW sector, for $A=8$ particles. Numbers below the lowest-lying
  states are overlaps between exact ground state and CF approximation, rounded to
  four digits. \label{compact8plot} }
\end{figure}

The yrast spectra for $N+M=8$ particles are shown in Fig. \ref{compact8plot} for $0 \leq L \leq N+M$. Each subplot gives the spectrum for a specific $(N,M)$ and thus a fixed pseudospin $\mathcal{S}_z\equiv(N-M)/2$. As mentioned before, pseudospin symmetry of the Hamiltonian makes it sufficient to study only the highest weight states $\mathcal{S}\equiv\mathcal{S}_z\equiv(N-M)/2$ in each spectrum. For example, the full spectrum of $(N,M)=(1,7)$ contains states of $\mathcal{S}_z\equiv3$ and $\mathcal{S}\equiv3,4$. Energy and overlap (with CF states) of a state $\left|n,\mathcal{S}=4,\mathcal{S}_z=3\right\rangle$ is identical to that of the HW state $S^+\left|n,\mathcal{S}=4,\mathcal{S}_z=3\right\rangle$ which is in the HW spectrum of $(N,M)=(0,8)$. Thus the full TI spectrum of (for example) $(N,M)=(2,6)$ should combine the HW spectrum of $(N,M)=(0,8),(1,7)$ and $(2,6)$.

For the angular momenta in the range $0\leq L <M$, the number of linearly independent CF states is equal to the number of basis states in the highest weight sector. Therefore, the diagonalization simply reproduces the exact spectrum. 
Even though these CF functions do not extract low energy states, the CF wave functions form a particularly convenient basis (in contrast to the bases of Slater permanents or elementary symmetric polynomials) because the CF states are by construction translationally invariant and highest weight eigenstates of $\mathcal{S}^2$.
For larger angular momenta $M\leq L \leq N+M$, the overlaps are less than unity, but still very high ($>0.99$) for the low lying states in the spectrum, and $>0.9$ for all but a very few cases. Since the dimension of the full CF space is only
slightly smaller the dimension of the complete Hilbert space, this is not very surprising.

TABLE \ref{dimtable} gives an overview of the dimensions of the spaces in the problem, along with the number of candidate CF wave functions, for $(N,M)=(2,6)$. Again we notice that the CF states span the whole sector for $L<M$. The number of distinct {\it candidate} (before projection) pairs of determinants tends to be considerably larger than the number of linearly independent CF basis states. We do not know of a systematic way to tell, {\it a priori}, which candidate determinants will produce identical (or linearly dependent) CF polynomials, and thus this has to be numerically checked explicitly.

\begin{table}[!htb]
{\renewcommand{\arraystretch}{1.5}
\renewcommand{\tabcolsep}{0.2cm}

 \begin{tabular}{|l|c|c|c|c|c|c|c|c|c|}
  \hline
  $L$ & 0 & 1 & 2 & 3 & 4 & 5 & 6 & 7 & 8 \\ \hline
  $d_H$ & 0 & 0 & 1 & 1 & 3 & 4 & 8 & 10 & 18 \\ \hline
  $d_{CF}$ & 0 & 0 & 1 & 1 & 3 & 4 & 7 & 8 & 16 \\ \hline
  $n_{CF}$ & 0 & 0 & 15 & 20 & 41 & 54 & 72 & 85 & 90 \\ \hline
  
 \end{tabular}
}
\caption{Dimensionality and number of compact CF candidates for $2+6$ particles. $d_H$ is the dimension of the TI-HW eigenspace, $d_{CF}$
is the number of linearly independent compact CF states, and $n_{CF}$ is the number of distinct candidate determinants. \label{dimtable}}
\end{table}
As seen above, the space of CF states does not extract any information about the low-lying states in particular. In addition, the large number of compact candidate determinants quickly becomes difficult to handle. As an example, the dimension of the TI-HW eigenspace at $(N,M)=(3,9)$ and $L=8$ is only 14, while the number of compact candidates is 1799, see TABLE \ref{numbertable}. We therefore seek to identify a subset of the CF states, to reduce their number without loosing the high overlaps with the lower energy exact states. The simple states have been found to accomplish just this.

\begin{table*}[!htb]
{\renewcommand{\arraystretch}{1.5}
\renewcommand{\tabcolsep}{0.2cm}

 \begin{tabular}{|l|c|c|c|c|c|c|c|c|c|c|c|c|c|}
  \hline
  $L$ & 0 & 1 & 2 & 3 & 4 & 5 & 6 & 7 & 8 & 9 & 10 & 11 & 12 \\ \hline
  compact & 0 & 0 & 0 & 84 & 168 & 441 & 772 & 1264 & 1799 & 2502 & 3022 & 2693 & 4310 \\ \hline
  simple & 0 & 0 & 0 & 84 & 168 & 266 & 326 & 379 & 392 & 407 & 388 & 373 & 318 \\ \hline
 \end{tabular}
}
\caption{Comparing the number of compact candidates to the number of simple candidates.
\label{numbertable}}
\end{table*}

\subsection{Simple CF diagonalization}
In TABLE \ref{numbertable} we compare, for different $L$, the number of candidate determinants for the compact states to the corresponding number of simple state-candidates at $(N,M)=(3,9)$. As we see, the latter is significally lower\cite{simplecomment}. At the two extremes $S_z=0$ and $S_z=A/2$ we have only simple candidates and no simple candidates, respectively. Because of the latter fact, and also the fact that systems with $N=0$ are not really two-component systems to begin with, we will omit such cases in the discussion.

\begin{figure}
  \centering
%%  \subfloat{\includegraphics[width=0.4\textwidth]{simple/0plus8nolab.pdf}}
%%  \quad
  \subfloat{\includegraphics[width=0.4\textwidth]{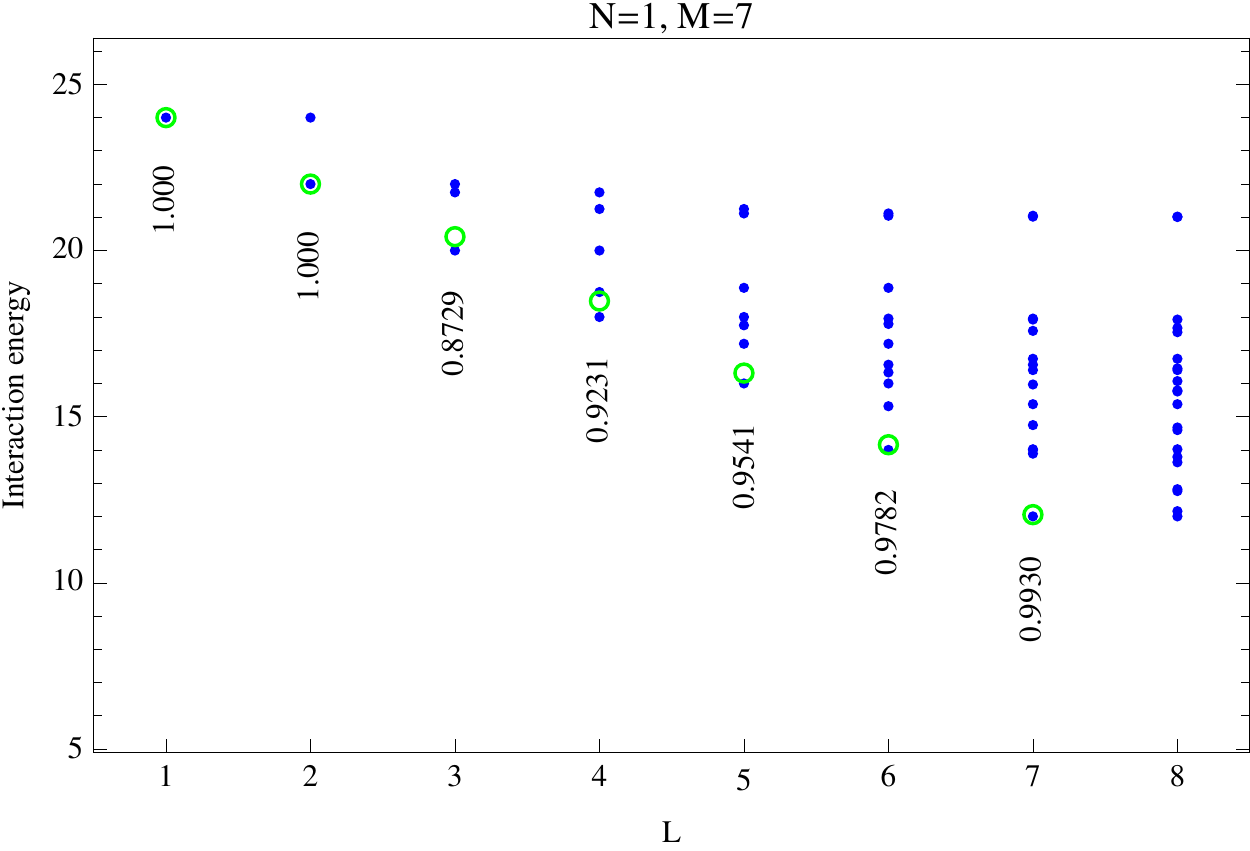}}
  \quad
  \subfloat{\includegraphics[width=0.4\textwidth]{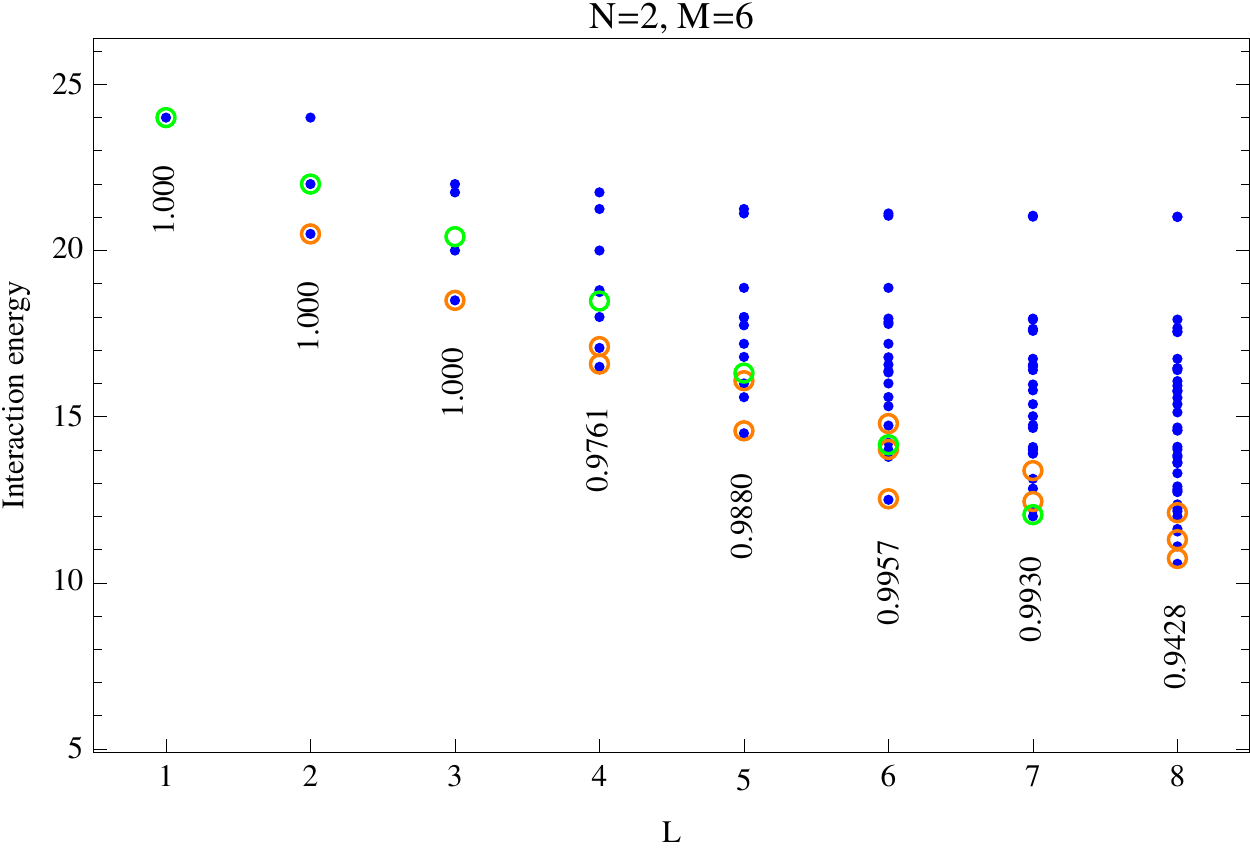}}
  \newline
  \subfloat{\includegraphics[width=0.4\textwidth]{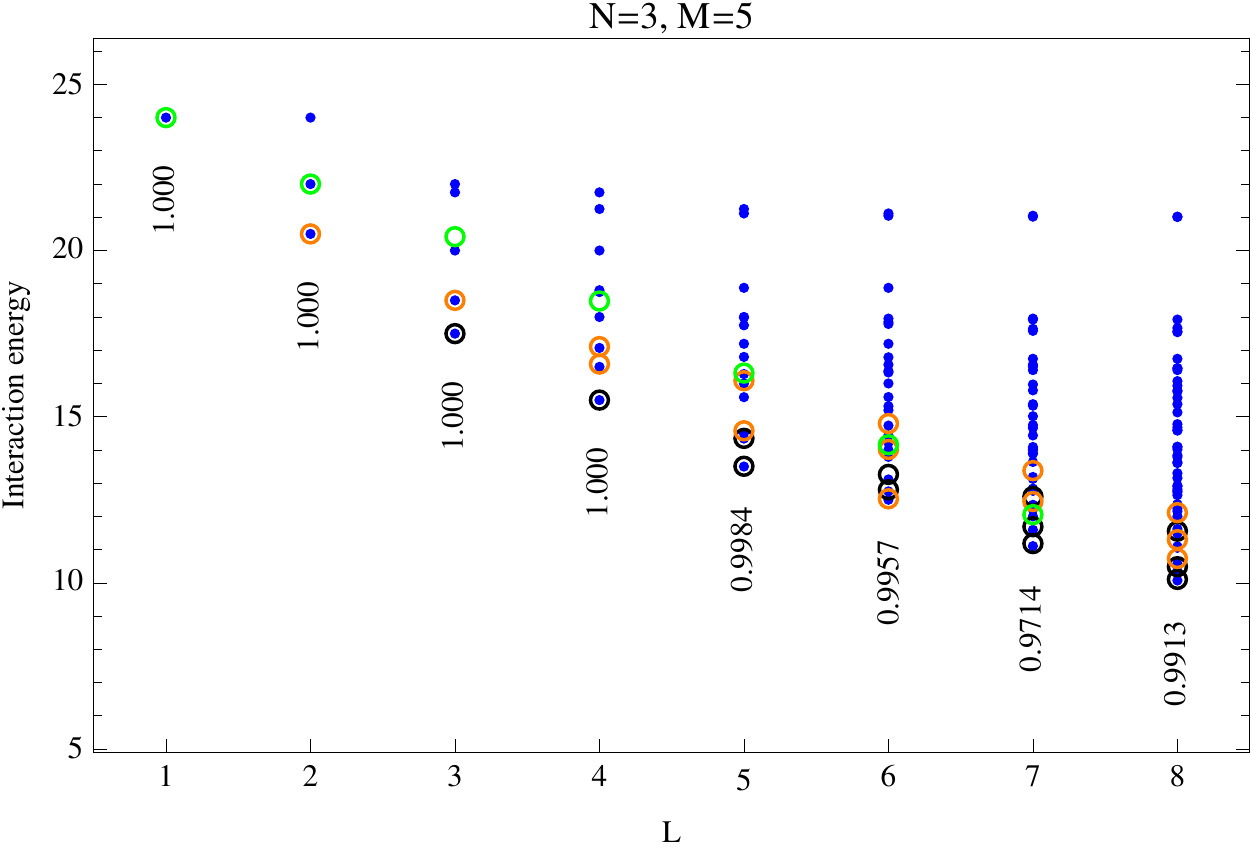}}
  \quad
  \subfloat{\includegraphics[width=0.4\textwidth]{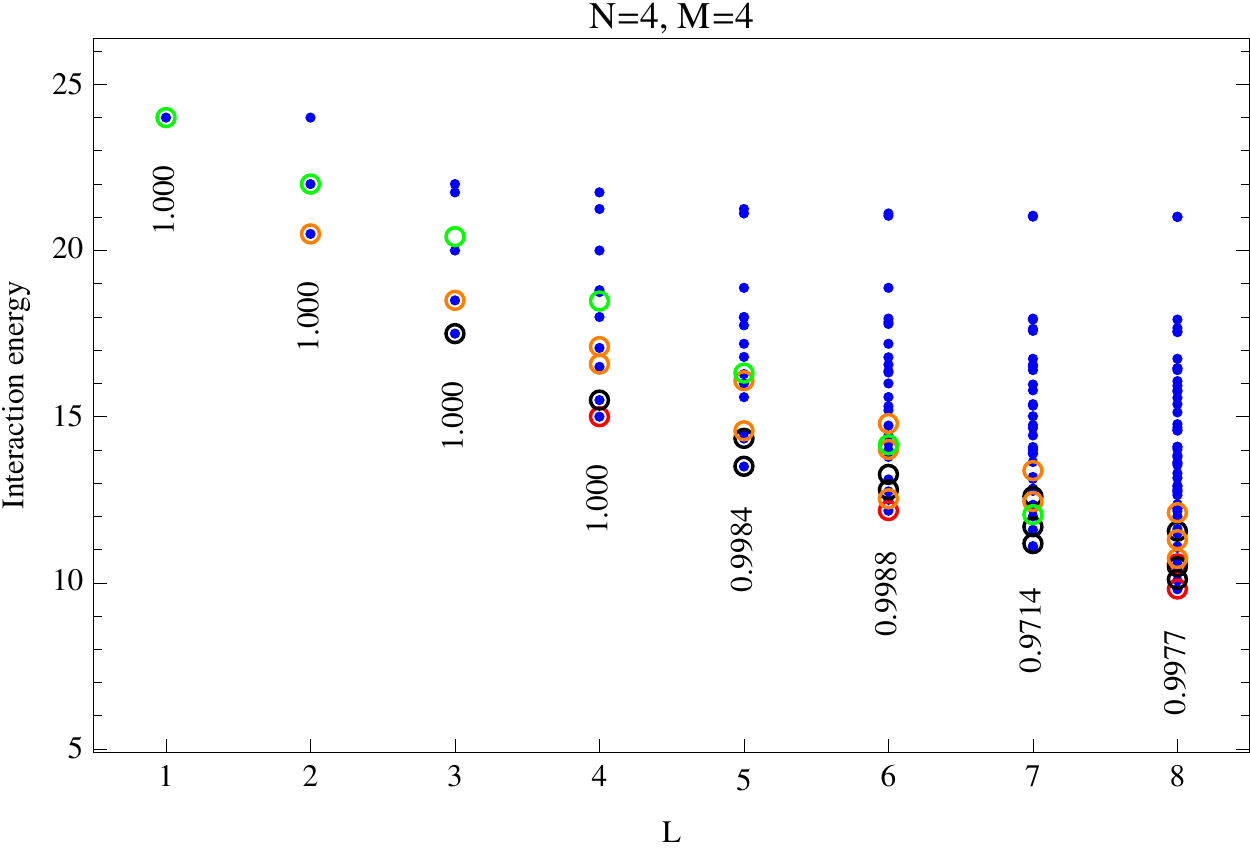}}
  \newline
  \caption{(Color online) Exact energy eigenstates (blue dots) and simple CF diagonalization
  results (colored rings) in the TI sector, for $A=8$ particles. All values of $S^2$ are included
  in this plot. The numbers denote overlaps between (lowest-lying) simple states and exact ground states. \label{simple8plot}}
\end{figure}

\begin{figure}
  \centering
  \subfloat{\includegraphics[width=0.4\textwidth]{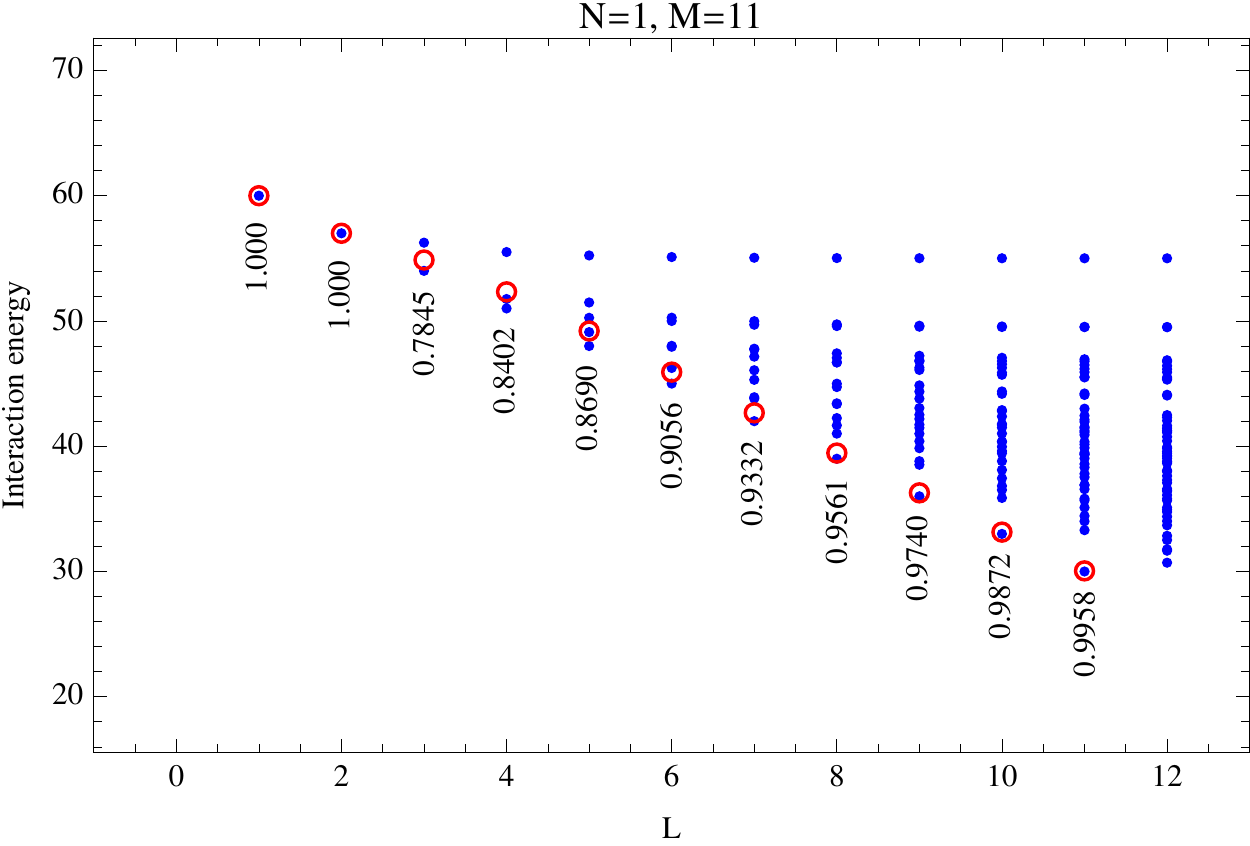}}
  \quad
  \subfloat{\includegraphics[width=0.4\textwidth]{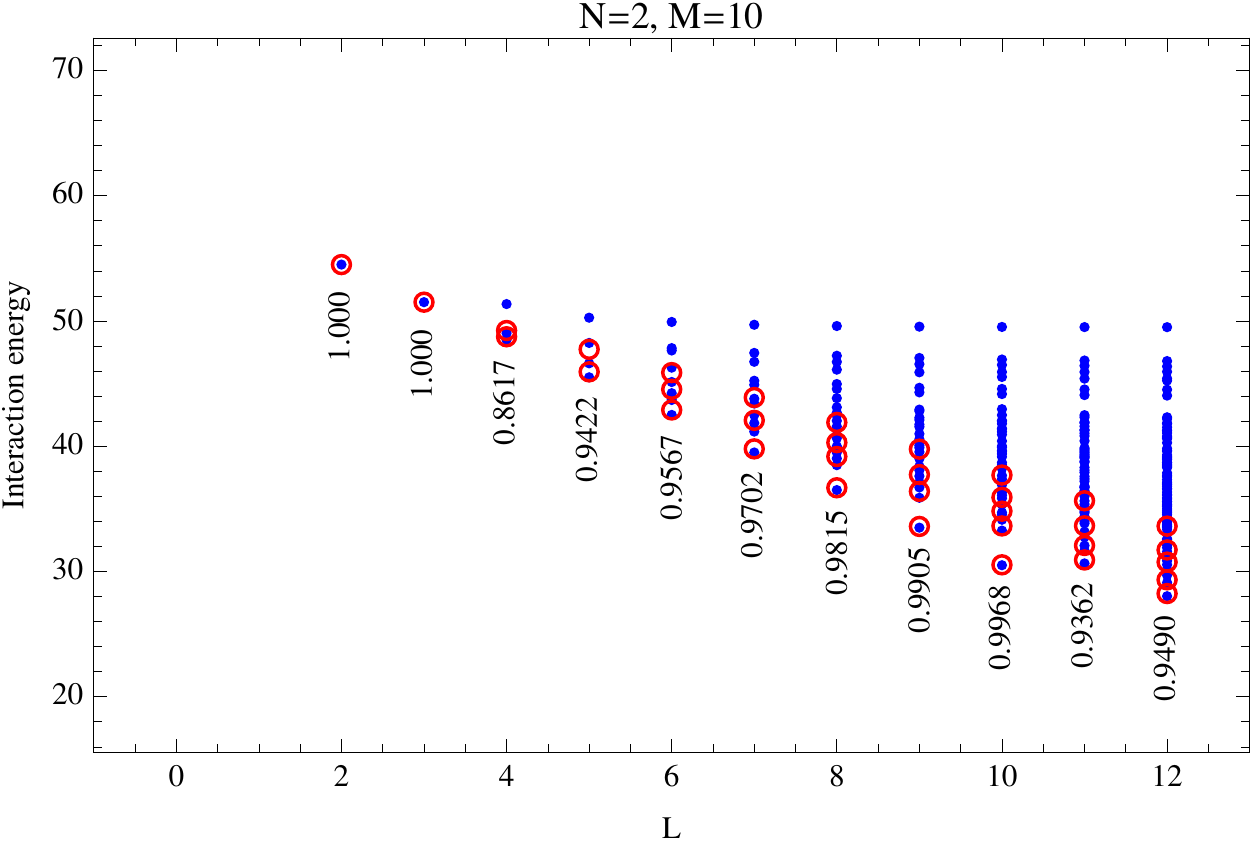}}
  \newline
  \subfloat{\includegraphics[width=0.4\textwidth]{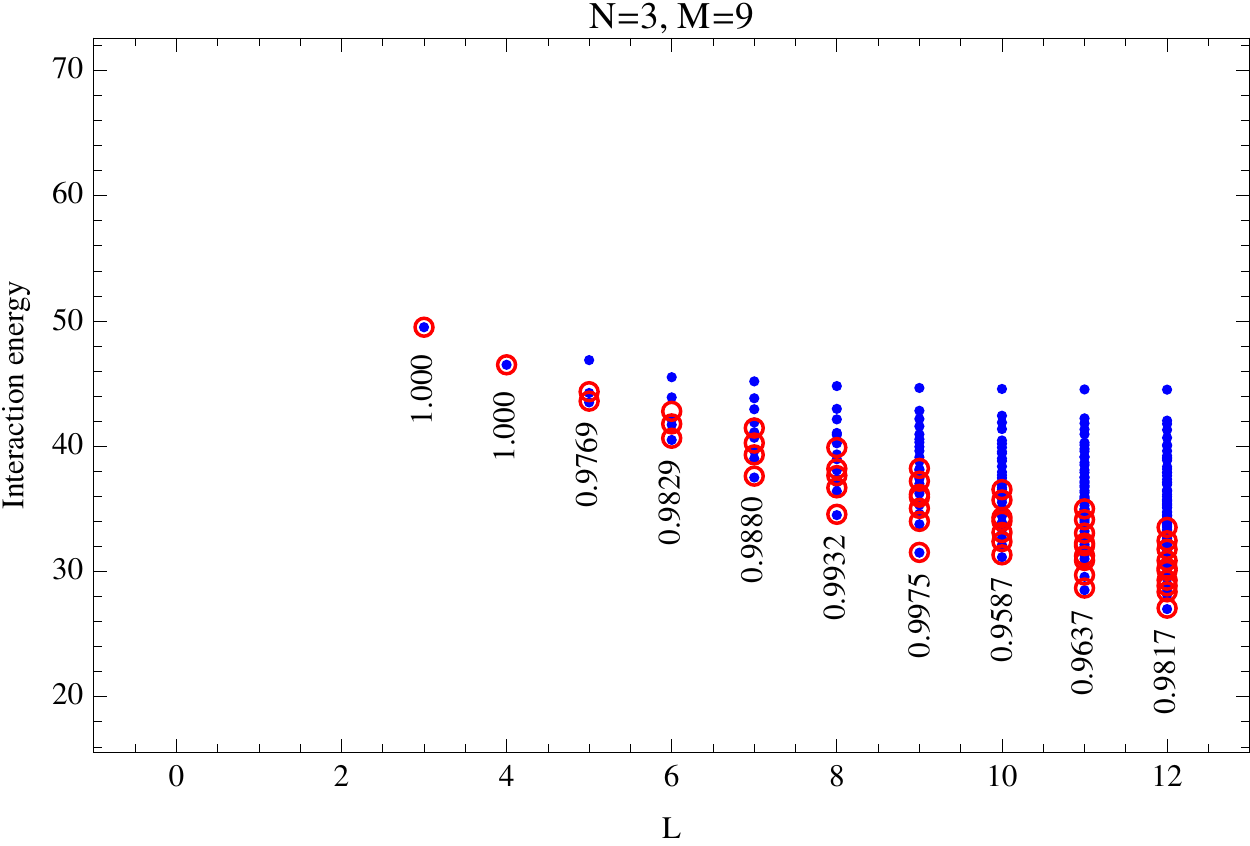}}
  \quad
  \subfloat{\includegraphics[width=0.4\textwidth]{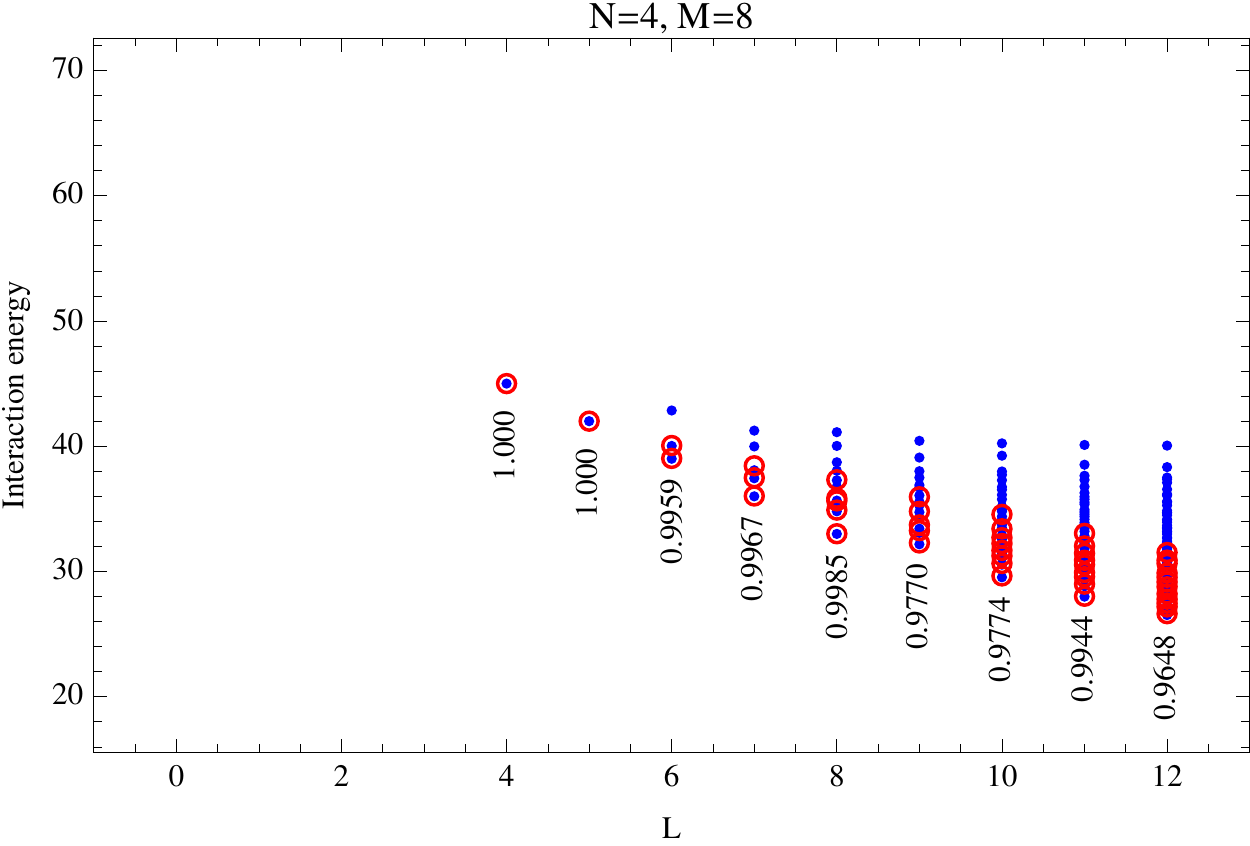}}
  \newline
  \subfloat{\includegraphics[width=0.4\textwidth]{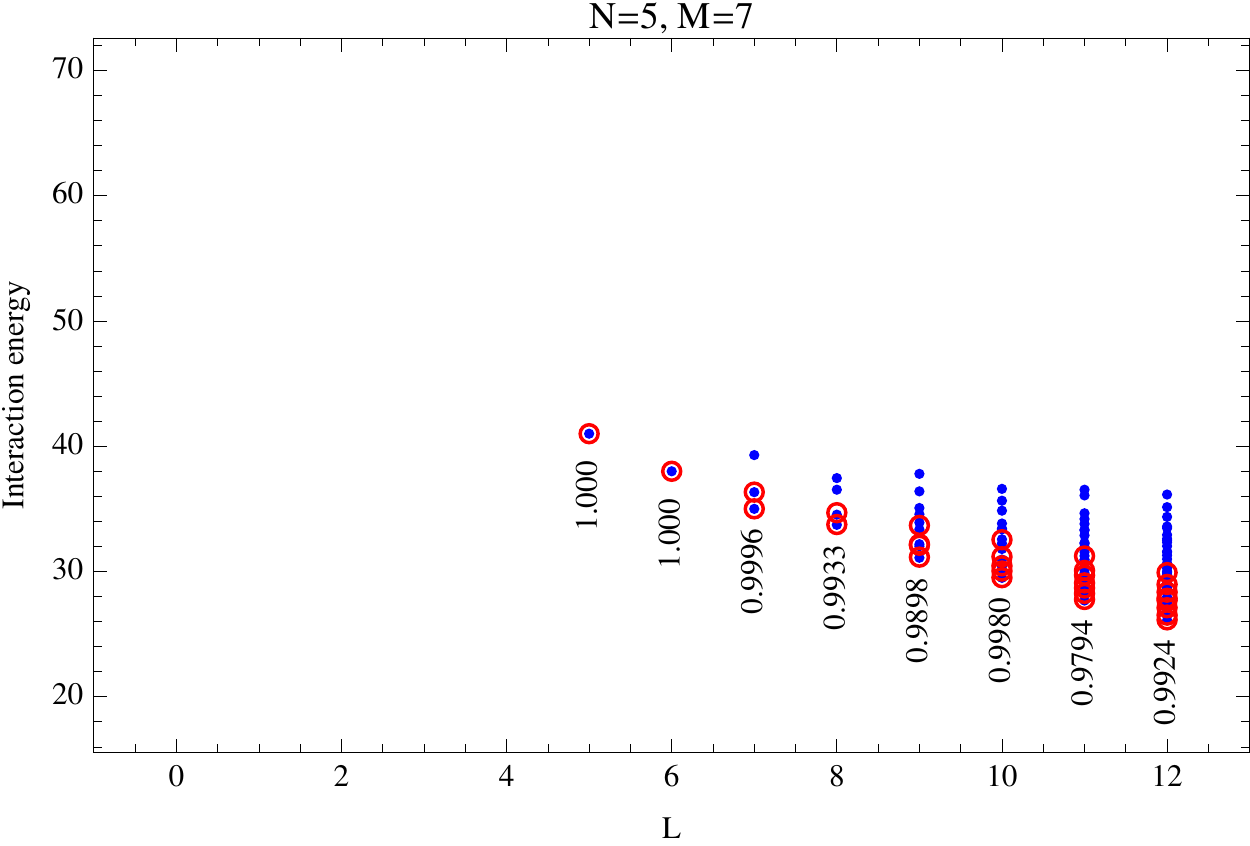}}
  \quad
  \subfloat{\includegraphics[width=0.4\textwidth]{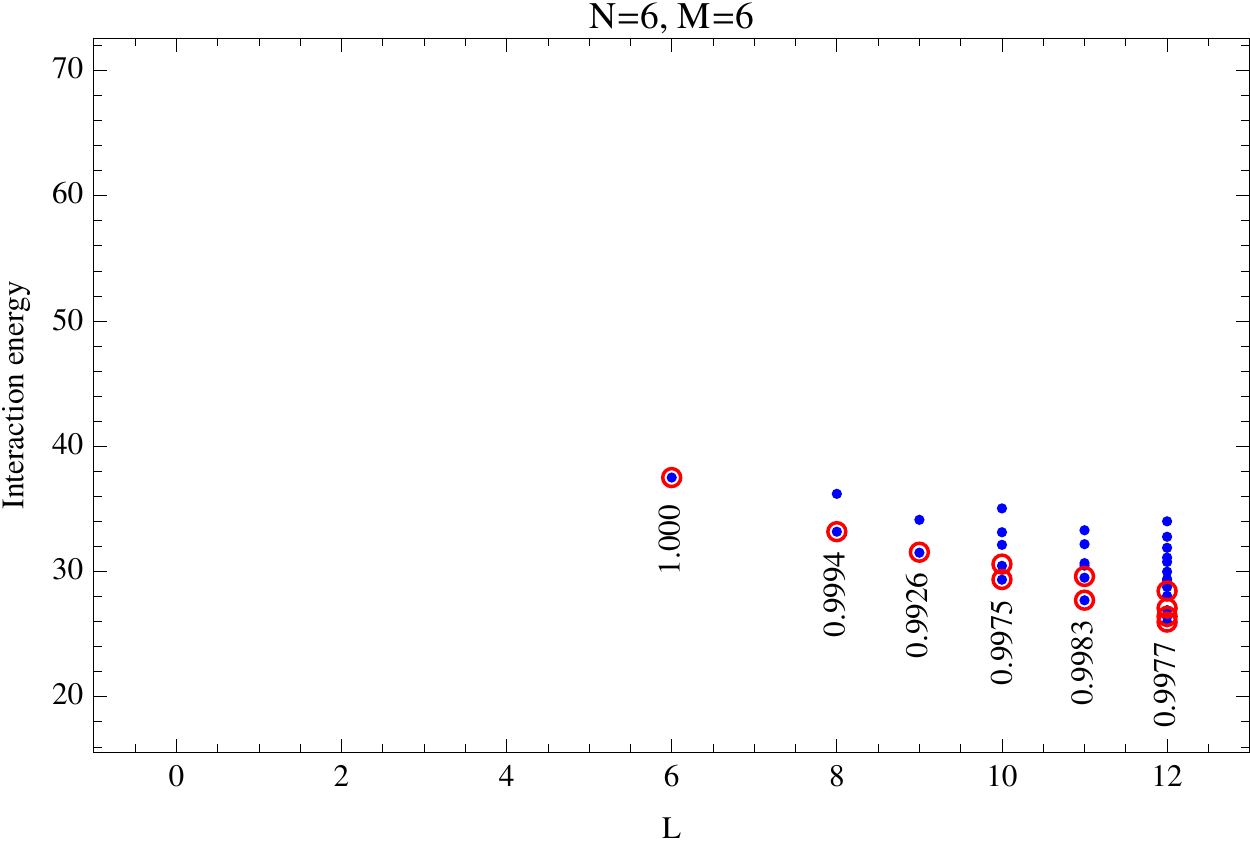}}
  \newline
  \caption{(Color online) Exact energy eigenstates (blue dots) and simple CF diagonalization
  results (red rings) in the TI-HW sector, for $A=12$ particles. The numbers denote overlaps between (lowest-lying) simple states and exact ground states. \label{simple12plot}}
\end{figure}

The yrast states are shown for 8 and 12 particles in FIGs \ref{simple8plot} and \ref{simple12plot}, this time with only the simple CF states included. Non-highest-weight states are also included in the spectrum. We clearly see that the simple states give very good approximations to the lowest-lying states, while the higher excitations are excluded. It is known \cite{papenbrock} that the ground states at angular momenta $N \leq L \leq M$ are highest weight states, while this is not necessarily the case for larger $L$. We find that the CF trial states give the correct pseudospin for the ground states. Generally, the simple states also cover most of the other low-lying spectrum with high accuracy.

The cases $(N,M)=(1,A-1)$ are especially interesting. As we see from FIGs. \ref{simple8plot} and \ref{simple12plot}, there is one unique CF wave function for each $L$, serving as ansatz states for the ground states. We also see that a gap to excitations opens up as $L$ approaches $L=A$ from below. For $L \geq A$ however, there is no gap. In CF language, this corresponds to a situation where the $\Lambda$-level ''kinetic'' energy decreases steadily as $L$ grows from 1 to $A-1$, forming simple states at
each $L$. It turns out, however, that there is no way to keep decreasing the $\Lambda$-level energy as $L$ increases from $A-1$ to $A$: at $A-1$ the state is
\begin{equation} \label{Nlik1}
\psi(\{ z_i\}, \{ w_i\}) = 
\begin{vmatrix}
\p^0_{z_1}
\end{vmatrix}
\cdot
\begin{vmatrix}
 \p^0_{w_1} & \p^0_{w_2} & \cdots & \p^0_{w_{A-1}}\\
\p^1_{w_1} & \p^1_{w_2} & \cdots & \p^1_{w_{A-1}} \\
\vdots & \vdots & \vdots & \vdots \\
\p^{A-1}_{w_1} & \p^{A-1}_{w_2} & \cdots & \p^{A-1}_{w_{A-1}}  \\
\end{vmatrix}
\, \prod_{k<l} (w_k - w_l) \prod_{n} (z_1 - w_n)
\end{equation} 
where the last part is the Jastrow factor for this system. There is no way to create a new simple state by increasing $L$ from the state (\ref{Nlik1}). Therefore there is no `simple' CF candidate for $L=A$. This matches well with an abrupt vanishing of the gap in exact spectrum, while going from $L=A-1$ to $A$.

\section{Discussion and outlook}
 \label{sec:concl}
In summary, we have shown that the CF prescription for writing ansatz wave functions for
the two-component system of rotating bosons reproduces the exact spectra with very high precision even at the lowest angular momenta (i.e. far outside the quantum Hall regime). This approach also provides a convenient way of keeping track of the good quantum numbers of the system. For the systems studied, we have shown that the compact CF states span the whole TI-HW sector of Hilbert space for $L < M$, and thus exactly reproduce the many-body spectra in each sector, giving a convenient basis with the desired properties. For larger $L$, we find overlaps remarkably close to unity. 

Secondly, we have identified the lowest-lying CF states for $N>0$ to be composed of states with lowest $\Lambda$-level kinetic energy(\ref{simple-spwf}). The restriction to this subset of CF states offers significant computational gains through reduction of dimensionality of the trial-function space while retaining very high overlaps with the low energy part of the exact eigenstates. The formalism that we have identified, thus provides very good set of wave functions that can be used to study low energy properties of the system. A possible future application is to use these to study details of the structure and formation of vortices in the system.

A very interesting future direction would be to go beyond the simple limit of homogeneous interactions, i.e. break the pseudospin symmetry that we have exploited in this paper. In particular, we plan to study the case where the interspecies interaction is tuned away from the intraspecies interaction (which is still assumed to be the same for both species). Interesting physics might be expected in this case, such as transitions from cusp states to non-cusps states on the yrast line, or other possibly experimentally verifiable phenomena. The particular case of $(N,M)=(1,A-1)$ is especially interesting due to presence of what appears to be a gapped mode that is captured well by the `simple' CF function. This case could be used to model the physics of an `impurity' boson on the rotating bose condensate.

Finally, it is worth commenting on the fact that the number of seemingly distinct CF candidate Slater determinants is much larger than the actual number of linearly independent CF wave functions. In particular, in many cases, several different CF Slater determinants turn out to produce the same final polynomial. This implies mathematical identities which, at this point, we fail to understand in a systematic way. This question links closely to a recent paper\cite{balram13} which discusses the apparent over-prediction of the number of CF states compared to the number of states seen in exact diagonalization, in excited bands of electronic quantum Hall states. Thus, it seems worthwhile to try and understand this issue in the general context of the composite fermion model.

%%%%%%%%%

\section*{Acknowledgement}
\noindent
We would like to thank Jainendra Jain and Thomas Papenbrock for very helpful discussions. This work was financially supported by the Research Council of Norway and by NORDITA.
\

\appendix

\section{Translational invariance}
\label{appA}

We show that an eigenstate $\Psi$ of the operator $L_c$ (see Eq. \ref{LC}) with eigenvalue 0 is
a translationally invariant state. Note that this translational invariance applies only to the polynomial part of the wavefunction and not the Gaussian part. Translation of a boson $z_j$ through a displacement $\vec c$ is achieved by the operator 
\begin{equation}
\hat{T}_j(\vec c) = e^{-i \vec c \cdot \hat p_j} = 1 -i \vec c \cdot \hat p_j - \frac{1}{2}{(\vec c \cdot \hat p}_j)^2 + \cdots
\end{equation}
While considering its action on wavefunctions in the lowest Landau level, we can set $\partial_{\bar{z}_j}$ to be $0$. Thus for any translation $c$, action of $\vec c \cdot \hat p_j$ is equivalent to $(c_x+\imath c_y)\partial_{z_j} \equiv c_z\partial_j$ in the lowest Landau level. Since $\p_i$ and $\p_j$ commute, the translation of all the $N+M$ particles in the system by the same vector is given by 
\begin{equation}
\begin{split}
\hat{T} & = \prod_{j=1}^{N+M} \hat{T}_j(\vec c) \\
& = \exp \left(-ic_z \sum_{j=1}^{N+M} \hat \p_j \right) \\
& = 1 - ic_z \sum_{j=1}^{N+M} \hat \p_j -\frac{c_z^2}{2} \left( \sum_{j=1}^{N+M} \hat \p_j \right)^2 + \cdots \\
\end{split}
\end{equation}
For an eigenstate of $L_c$ with eigenvalue 0,
\begin{equation}
L_c \Psi(z,w) = R \left( \sum_{j=1}^{N+M} \partial_j \right) \Psi(z,w) = 0
\end{equation}
which implies $\left( \sum_{j=1}^{N+M} \partial_j \right) \Psi(z,w) = 0$. Therefore, we conclude
\begin{equation}
\begin{split}
\hat{T} \Psi(z,w) & = \Psi(z,w) -ic_z \left( \sum_{j=1}^{N+M} \partial_j \right) \Psi(z,w) - \frac{c_z^2}{2}\left( \sum_{j=1}^{N+M} \partial_j \right)^2 \Psi(z,w) + \cdots \\
& = \Psi(z,w)
\end{split}
\end{equation}
That is, $\Psi(z,w)$ is translationally invariant.

\section{Numerical Methods}
\label{appB}
\subsection*{Exact Spectrum}
The exact spectrum was obtained by diagonalizing the model Hamiltonian in the subspace of translationally invariant and highest weight pseudospin states  (TI-HW states). The TI-HW states are obtained by finding the null space for $\mathcal{O}=\mathcal{S}_- \mathcal{S}_+ + L_c$ where $\mathcal{S}_+$ and $\mathcal{S}_-$ are the pseudo spin raising and lowering operators respectively, and $L_c$ is the center-of-mass angular momentum operator, as before. Since $\mathcal{S}_-\mathcal{S}_+$ and $L_c$ are positive definite, $\left\langle L_c \right\rangle=0$ and $\left\langle \mathcal{S}_- \mathcal{S}_+\right\rangle=0$ is equivalent to $\left\langle \mathcal{O} \right\rangle=0$. Exact diagonalization was performed using Lanczos algorithm.

\subsection*{Projection of CF states}

In this section, we summarize the idea behind the algorithm used for projection calculation. The actual implementation uses representation of monomials of $z$ and $w$ in the computer as integer arrays. We use the symbol $\p^n_i$ for $n^{\rm th}$ derivatives with respect to $z_i$ and $D^n_i$ for derivatives with respect to $w_i$. The projection operation, in the case of compact states, involves evaluation of action of complicated derivatives on the Jastrow factor, of the following form
\begin{equation}
\Psi=\det\left[\begin{array}{ccc}
\partial_{1}^{n_{1} }z_{1}^{m_{1}} & \partial_{2}^{n_{1} }z_{2}^{m_{1}} & \dots\\
\partial_{1}^{n_{2}} z_{1}^{m_{2}} & \partial_{2}^{n_{2}} z_{2}^{m_{2}}& \dots\\
\vdots & \vdots & \ddots
\end{array}\right]\times\det\left[\begin{array}{ccc}
D_{1}^{\overline{n}_{1}} w_1^{\overline{m}_{1}} & D_{2}^{\overline{n}_{1}} w_{2}^{\overline{m}_{1}} & \dots\\
D_{1}^{\overline{n}_{2}} w_{1}^{\overline{m}_{2}} & D_{2}^{\overline{n}_{2}} w_{2}^{\overline{m}_{2}} & \dots\\
\vdots & \vdots & \ddots
\end{array}\right]J(z,w)
\label{beforeproj}
\end{equation}

Note that the derivatives $\p_i$ and $D_i$ act also on the Jastrow factor. 

Since $\Psi$ is symmetric in $\{z_1,z_2,\dots z_{N}\}$ and in $\{w_1,\dots w_{M}\}$, it has an expansion $\Psi=\sum C_{\lambda,\mu} \mathcal{M}_{\lambda,\mu}$ in the following basis functions
\begin{equation}
\mathcal{M}_{\lambda,\mu}=\mathcal{N}_{\lambda,\mu}\times \mathrm{sym}\left[ z^{\lambda_1}_1 z^{\lambda_2}_2 \dots z^{\lambda_{N}}_{N}\right]
\times
\mathrm{sym}\left[ w^{\mu_1}_{1} w^{\mu_2}_{2} \dots w^{\mu_{M}}_{M}\right]
\label{projbasis}
\end{equation}
where $\mathcal{N}_{\lambda,\mu}$ is the normalization. Partitions $\lambda$ and $\mu$ of length $N$ and $M$ index the basis states. These functions form a poor choice of basis functions for expanding translation invariant pseudospin eigenstates. However they are convenient for computational manipulations as they form an orthonormal basis set and can be represented easily on a computer in the form of sorted integer arrays $\lambda$ and $\mu$. The coefficients  $C_{\lambda,\mu}$ of the expansion can be obtained from computing the coefficients of expansion in a slightly simpler problem as summarized below. 

\subsubsection*{Simpler problem}

When the determinants in Eq. \ref{beforeproj} are expanded in permutations of the matrix elements, we get $\Psi=\sum_{P\in s_{N}} \sum_{Q\in s_{M}} \phi_{P,Q}$ where $s_{K}$ is the set of permutations of $\{1,2,3..K\}$ and $\phi_{P,Q}$ is

\begin{equation}
\phi_{P,Q}=\mathcal{O}_{P,Q}J(z,w)\\
\label{minimal-terms}
\end{equation}
where $\mathcal{O}_{P,Q}$ is given by
\begin{equation}
\mathcal{O}_{P,Q}=\left[(-1)^P \prod_{i=1}^{N} \partial_{P(i)}^{n_i} z^{m_i}_{P(i)} \right]
\left[(-1)^Q \prod_{j=1}^{M} \partial_{Q(j)+N}^{\bar{n}_j}  z^{\bar{m}_j}_{Q(j)+N}\right],\nonumber
\end{equation}
and the Jastrow factor has the expansion:
\begin{equation}
J(z,w)=\sum_{R\in s_{N+M}} (-1)^R \prod_{k=1}^{N+M} z_k^{R(k)}.\nonumber
\end{equation}
where $z_k=w_{k-N}$ for $k=N+1,N+2\dots N+M$. 

Consider the orthogonal basis functions containing monomials: $\mathfrak{m}_\lambda=\prod_{k=1}^{N+M}z_k^{\lambda_k}$. Here $\lambda$ is any non-negative integer sequence of length $N+M$. The Jastrow factor can be easily expanded in this basis as $J(z,w)=\sum_\lambda J_\lambda \mathfrak{m}_\lambda$, where $J_\lambda=1$ when $\lambda$ is an even permutation of $(0,1,2,3\dots N+M-1)$ and $-1$ for odd permutations.

All we need to calculate is $\phi_{P=\mathbb{1},Q=\mathbb{1}}$. Action of $\mathcal{O}_{\mathbb{1},\mathbb{1}}$ on $\mathfrak{m}_\lambda$ gives either $0$ or
\begin{eqnarray}
\mathcal{O}_{\mathbb{1},\mathbb{1}} \mathfrak{m}_\lambda &=& \mathfrak{m}_\mu \prod_{i=1}^{N+M}\frac{(\lambda_i+m_i) !}{(\lambda_i+m_i-n_i)!}\\
\mu_i &=& \lambda_i+m_i-n_i\nonumber \text{ where } i=1,\dots N+M
\end{eqnarray}
$\mathcal{O}_{\mathbb{1},\mathbb{1}} \mathfrak{m}_\lambda=0$ if $\mu_i<0$ for any $i$. Since we know $J_\lambda$ and the action of $\mathcal{O}$ on the basis functions, we can evaluate $\phi_{\mathbb{1},\mathbb{1}}$ to get the coefficients $c_\lambda$ in the expansion of the form:
\begin{equation}
\phi_{\mathbb{1},\mathbb{1}}=\sum c_{\lambda} \mathfrak{m}_\lambda
\end{equation}
The above coefficients $c_\lambda,\mu$ are related to the coefficients $C_{\lambda,\mu}$ as shown below.

From, Eq.\ref{minimal-terms} it can be seen that $\phi_{P,Q}(z,w)=\phi_{\mathbb{1},\mathbb{1}}(Pz,Qw)$ due to the antisymmetry of $J$. From the expansion of $\Psi$ in  $\phi_{P,Q}$, we have 
\begin{equation}
\Psi=\sum_\lambda c_\lambda \sum_{P\in s_N} \sum_{Q\in s_M} \mathfrak{m}_\lambda(Pz,Qw)
\label{exp2}
\end{equation}

The array of exponents $\lambda$ can be split into exponents $\lambda^z\equiv(\lambda_1\dots\lambda_N)$ of $z_k$ and exponents $\lambda^w\equiv(\lambda_{N+1}\dots\lambda_{N+M})$ of $w_k$ ($=z_{k+N}$). Represent the sorted form of an array $\lambda^{z(w)}$ as $\tilde{\lambda}^{z(w)}$. In this notation,
\begin{equation}
\sum_{P\in s_N} \sum_{Q\in s_M} \mathfrak{m}_\lambda(Pz,Qw) = \frac{1}{\mathcal{N}_{\tilde{\lambda}^z,\tilde{\lambda}^w}}\mathcal{M}_{\tilde{\lambda}^z,\tilde{\lambda}^w}
\end{equation}
Comparing the expansion Eq. \ref{exp2} to the expansion of $\Psi$ in terms of $\mathcal{M}$, we get the relation that can be used to calculate
\begin{equation}
C_{\alpha,\beta}=\sum_{\lambda}' \frac{c_\lambda}{\mathcal{N}_{\tilde{\lambda}^z,\tilde{\lambda}^w}}
\end{equation}
where the sum is over all $\lambda$, such that the parts $\tilde{\lambda}^z$ equals $\alpha$ and $\tilde{\lambda}^w$ equals $\beta$.
\vskip 3mm\noi


\begin{thebibliography}{99}
\bibitem{reviews} S. Viefers, J. Phys.: Cond. Mat. {\bf 20}, 123202 (2008); N. Cooper, Advances in Physics {\bf 57}, 539 (2008).
\bibitem{roncaglia11} M. Roncaglia, M. Rizzi, and J. Dalibard, www.nature.com, Scientific Reports {\bf 1}, doi:10.1038/srep00043 (2011).
\bibitem{lin09} Y.-J. Lin, R. L. Compton, K. Jim\'enez-Garcia, J. V. Porto, and I. B. Spielman, Nature {\bf 462}, 628 (2009).
\bibitem{dalibardreview} J. Dalibard, F. Gerbier, G. Juzeliunas, and P. \"Ohberg, Rev. Mod. Phys. {\bf 83}, 1523 (2011).
\bibitem{julia-diaz} B. Julia-Diaz, T. Grass, N. Barberan, M. Lewenstein, New J. Phys. {\bf 14}, 055003 (2012).


\bibitem{mottelson99} B. Mottelson, Phys. Rev. Lett. {\bf 83}, 2695 (1999).
\bibitem{bertsch99} G. Bertsch and T. Papenbrock, Phys. Rev. Lett. {\bf 83}, 5412 (1999).
\bibitem{smith00} R. A. Smith and N. K. Wilkin, Phys. Rev. A {\bf 62}, 061602 (2000).
\bibitem{kavoulak01} A. D. Jackson, G. M. Kavoulakis, B. Mottelson, and S. M. Reimann, Phys. Rev. Lett. {\bf 86}, 945 (2001).
\bibitem{korslund} N. Korslund and S. Viefers, Phys. Rev. A {\bf 73}, 063602 (2006).
\bibitem{viefers10} S. Viefers and M. Taillefumier, J. Phys. B {\bf 43}, 155302 (2010).

\bibitem{gemelke10} N. Gemelke, E. Sarajlic, and S. Chu, arXiv:1007.2677.



\bibitem{mondugno02} G. Modugno, M. Modugno, F. Riboli, G. Roati, and M. Inguscio, Phys. Rev. Lett. {\bf 89}, 190404 (2002).
\bibitem{bloch01} I. Bloch, M. Greiner, O. Mandel, T. W. H�nsch, and T. Esslinger, Phys. Rev. A {\bf 64}, 021402(R) (2001). 
\bibitem{hall98} D. S. Hall, M. R. Matthews, J. R. Ensher, C. E. Wieman, and E. A. Cornell, Phys. Rev. Lett. {\bf 81}, 1539 (1998);
V. Schweikhard, I. Coddington, P. Engels, S. Tung, and and E. A. Cornell, Phys. Rev. Lett. {\bf 93}, 210403 (2004).

\bibitem{papp08} S. B. Papp, J. M. Pino, and C. E. Wieman, Phys. Rev. Lett. {\bf 101}, 040402 (2008).
\bibitem{grass12} T. Grass, B. Julia-Diaz, N. Barberan, and M. Lewenstein, Phys. Rev. A {\bf 86}, 021603(R) (2012).
\bibitem{ueda12} S. Furukawa and M. Ueda, Phys. Rev. A {\bf 86}, 031604(R) (2012).
\bibitem{jain13} Y.-H. Wu and J. K. Jain, Phys. Rev. B {\bf 87}, 245123 (2013).
\bibitem{furukawa13} S Furukawa, and M. Ueda, Phys. Rev. Lett. {\bf 111}, 090401 (2013).
\bibitem{senthil13} T. Senthil, and M. Levin, Phys. Rev. Lett. {\bf 110}, 046801 (2013).
\bibitem{grass13} T. Grass, D. Raventos, M. Lewenstein, and B. Julia-Diaz, arXiv:1310.3709.

\bibitem{bargi07} S. Bargi, J. Christensson, G. M. Kavoulakis, and S. M. Reimann, Phys. Rev. Lett. {\bf 98}, 130403 (2007).
\bibitem{papenbrock} T. Papenbrock, S. M. Reimann, and G. M. Kavoulakis, Phys. Rev. Lett. {\bf 108}, 075304 (2012).
\bibitem{jainbook} J. K. Jain, {\it Composite Fermions}, Cambridge University Press (2007).
\bibitem{viefers00} S. Viefers, T. H. Hansson, and S. M. Reimann, Phys. Rev. A {\bf 62}, 053604 (2000).


\bibitem{grasscomment} A preprint\cite{grass13} that appeared during the completion of this work, also uses the CF approach in the disk geometry, though with focus on the higher-angular momentum incompressible states and their edge excitations. 

\bibitem{simplecomment} Again, the number of simple candidates is, in general, larger than the number of linearly independent simple states.

%\bibitem{mariusmaster} M. L. Meyer, Master Thesis, University of Oslo (2013).
\bibitem{balram13} A. C. Balram, A. Wojs, and J. K. Jain, Phys. Rev. B {\bf 88}, 205312 (2013).
\end{thebibliography}
\end{document}